\documentclass[5p,times,twocolumn]{elsarticle}
\usepackage{physics}
\usepackage{tabularx}
\usepackage{siunitx}
\usepackage{xcolor}
\usepackage{hyperref}
\hypersetup{hidelinks}
\usepackage{csquotes}
\usepackage{amssymb}% http://ctan.org/pkg/amssymb
\usepackage{pifont}% http://ctan.org/pkg/pifont
\newcommand{\cmark}{\ding{51}}%
\newcommand{\xmark}{\ding{55}}%
\usepackage{acro}
\DeclareAcronym{DR}{
    short = DR,
    long  = Demand Response,
}
\DeclareAcronym{OB}{
    short = OB,
    long  = Occupant Behavior,
}
\DeclareAcronym{HVAC}{
 short = HVAC,
 long  = {Heating, Ventilation, and Air Conditioning},
}
\DeclareAcronym{PV}{
 short = PV,
 long  = {Photovoltaic},
}
\DeclareAcronym{MPC}{
    short = MPC,
    long  = Model Predictive Control,
}
\DeclareAcronym{PI}{
    short = PI,
    long  = Proportional Integral,
}
\DeclareAcronym{RC}{
    short = RC,
    long  = Resistor Capacitor,
}
\DeclareAcronym{PMV}{
    short = PMV,
    long  = Predicted Mean Vote,
}
\DeclareAcronym{PPD}{
    short = PPD,
    long  = Predicted Percentage of Dissatisfied,
}
\DeclareAcronym{OTS}{
    short = OTS,
    long  = Occupants' Thermal Satisfaction,
}
\DeclareAcronym{KPIs}{
    short = KPIs,
    long  = Key Performance Indicators,
}
\DeclareAcronym{RBC}{
    short = RBC,
    long  = Rule-based Control,
}
\DeclareAcronym{PSC}{
    short = PSC,
    long  = Price Storage Control,
}
\DeclareAcronym{PC}{
    short = PC,
    long  = Price Control,
}
\DeclareAcronym{EU}{
    short = EU,
    long  = European Union,
}

\graphicspath{{Figures/}}
\usepackage{lipsum}
\usepackage{booktabs}
\usepackage{float}
\usepackage{comment}
\usepackage{subcaption}
\usepackage{framed} % Framing content
\usepackage{multicol} % Multiple columns environment
\usepackage{nomencl} % Nomenclature package

\journal{Applied Energy}

\begin{document}

\begin{frontmatter}

\title{Occupant-Oriented Demand Response \\with Multi-Zone Thermal Building Control}

\author[1]{Moritz Frahm}
\author[2]{Thomas Dengiz}
\author[1]{Philipp Zwickel}
\author[1]{Heiko Maaß}
\author[1]{Jörg Matthes}
\author[1]{Veit Hagenmeyer}
\address[1]{Karlsruhe Institute of Technology, Institute for Automation and Applied Informatics, Eggenstein-Leopoldshafen, Germany}
\address[2]{Karlsruhe Institute of Technology, Institute for Industrial Production, Karlsruhe, Germany}

\begin{abstract}
\acresetall
 In future energy systems with high shares of renewable energy sources, the electricity demand of buildings has to react to the fluctuating electricity generation in view of stability. As buildings consume one-third of global energy and almost half of this energy accounts for \ac{HVAC} systems, \ac{HVAC}
 are suitable for shifting their electricity consumption in time. To this end, intelligent control strategies are necessary as the conventional control of HVAC is not optimized for the actual demand of occupants and the current situation in the electricity grid. In this paper, we present the novel multi-zone controller \ac{PSC} that not only considers room-individual \ac{OTS}, but also the available energy storage, and energy prices. The main feature of \ac{PSC} is that it does not need a building model or forecasts of future demands to derive the control actions for multiple rooms in a building. For comparison, we use an ideal, error-free \ac{MPC}, a heuristic control approach from the literature (PC), and a conventional hysteresis-based two-point control as upper and lower benchmarks. We evaluate the four controllers in a multi-zone environment for heating a building in winter and consider two different scenarios that differ in how much the permitted temperatures vary. 
 In addition, we compare the impact of model parameters with high and low thermal capacitance.
 The results show that \ac{PSC} strongly outperforms the conventional control approach in both scenarios and for both parameters concerning the electricity costs and \ac{OTS}. For high capacitance, it leads to \SI{22}{\percent} costs reduction while the ideal \ac{MPC} achieves cost reductions of more than \SI{39}{\percent}. Considering that \ac{PSC} does not need any building model or forecast, as opposed to \ac{MPC}, the results support the suitability of our developed control strategy for controlling \ac{HVAC} systems in future energy systems.

\end{abstract}

\begin{comment}
\begin{graphicalabstract}
\includegraphics{grabs}
\end{graphicalabstract}

\begin{highlights}
\item Research highlight 1
\item Research highlight 2
\end{highlights}
\end{comment}


\end{frontmatter}

\setlength{\nomitemsep}{-\parskip} % Baseline skip between items
\makenomenclature
\renewcommand*\nompreamble{\begin{multicols}{2}}
\renewcommand*\nompostamble{\end{multicols}}

\ExplSyntaxOn
\NewExpandableDocumentCommand{\strcase}{mm}
 {
  \str_case:nn { #1 } { #2 }
 }
\ExplSyntaxOff

\renewcommand\nomgroup[1]{%
  \item[\bfseries
    \strcase{#1}{
      {P}{Parameters}
      {A}{Acronyms}
      {V}{Variables}
      % add here other cases
    }%
  ]%
}

\begin{table*}[htb]
  \begin{framed}
  %Acronyms
    \nomenclature[A]{\acs{DR}}{\acl{DR}}
    \nomenclature[A]{\acs{HVAC}}{\acl{HVAC}}
    \nomenclature[A]{\acs{KPIs}}{\acl{KPIs}}
    \nomenclature[A]{\acs{MPC}}{\acl{MPC}}
    \nomenclature[A]{\acs{OB}}{\acl{OB}}
    \nomenclature[A]{\acs{OTS}}{\acl{OTS}}
    \nomenclature[A]{\acs{PV}}{\acl{PV}}
    \nomenclature[A]{\acs{PMV}}{\acl{PMV}}
    \nomenclature[A]{\acs{PPD}}{\acl{PPD}}
    \nomenclature[A]{\acs{PSC}}{\acl{PSC}}
    \nomenclature[A]{\acs{PC}}{\acl{PC}}
    \nomenclature[A]{\acs{RBC}}{\acl{RBC}}
    \nomenclature[A]{\acs{RC}}{\acl{RC}}
    \nomenclature[A]{\acs{PI}}{\acl{PI}}

%
  % Variables
    \nomenclature[V]{$T_\text{a}$}{ambient temperature in \si{\celsius}}
    \nomenclature[V]{$\dot{q}_{\mathrm{s}}$}{solar radiation in \si{\watt \per \square \metre}}
    \nomenclature[V]{$\dot{Q}_{\mathrm{h}_j}$}{heat flow of heat pump in \si{\watt}}
    \nomenclature[V]{$T_{\mathrm{i}_j}$}{room air temperature in \si{\celsius}}
    \nomenclature[V]{$T_{\mathrm{m}_j}$}{heat accumulating medium temperature in \si{\celsius}} 
    \nomenclature[V]{$\widehat{F}$}{empirical distribution function}
    \nomenclature[V]{$t$}{time in \si{\second}}
    \nomenclature[V]{$P_\text{el}$}{electrical power of heat pump in \si{\watt}}
    \nomenclature[V]{$\varepsilon _\text{h}$}{coefficient of performance of heat pump}
    \nomenclature[V]{$\chi_{\text{s}_j}$}{storage factor}
    \nomenclature[V]{$\chi_\text{p}$}{price factor}
    \nomenclature[V]{$\chi_\text{mod}$}{heat pump modulation degree}
    \nomenclature[V]{$\chi_\text{dis}$}{discomfort factor}
    \nomenclature[V]{$S_j$}{state of thermal charge}
    \nomenclature[V]{$P_\text{max}$}{max. electrical power of heat pump in \si{\watt}}
    \nomenclature[V]{$P_\text{buy}$}{bought electrical power in \si{\watt}}
    \nomenclature[V]{$U$}{electric voltage in \si{\volt}}
    \nomenclature[V]{$I$}{electric current in \si{\ampere}}
    \nomenclature[V]{$u$}{control input}
    \nomenclature[V]{$y$}{control output}
    \nomenclature[V]{$x$}{state}
    \nomenclature[V]{$p_\mathrm{tv}$}{time-variable parameter}
    \nomenclature[V]{$T$}{temperature in \si{\celsius}}
    \nomenclature[V]{$\dot{Q}$}{heat flow in \si{\watt}}
%
  % Parameters
    \nomenclature[P]{$g_{\mathrm{s}_j}$}{solar heat gain factor in \si{\square \metre}}
    \nomenclature[P]{$C_{\mathrm{i}_j}$}{heat capacity of room air in \si{\joule \per \kelvin}}
    \nomenclature[P]{$C_{\mathrm{m}_j}$}{heat capacity of heat accumulating medium in \si{\joule \per \kelvin}}
    \nomenclature[P]{$R_{\mathrm{i}_j}$}{resistance between $T_{\mathrm{i}_j}$ and $T_{\mathrm{m}_j}$ in \si{\kelvin \per \watt}}
    \nomenclature[P]{$R_{\mathrm{a}_j}$}{resistance between  $T_{\mathrm{i}_j}$ and $T_\mathrm{a}$ in \si{\kelvin \per \watt}}
    \nomenclature[P]{$T_{\mathrm{lb}_j}$}{minimal comfort temperature in \si{\celsius}}
    \nomenclature[P]{$T_{\mathrm{ub}_j}$}{maximal comfort temperature in \si{\celsius}}
    \nomenclature[P]{$T_{\mathrm{r}_j}$}{reference comfort temperature in \si{\celsius}}  
    \nomenclature[P]{$\Delta t_k$}{time step in \si{\second}}
    \nomenclature[P]{$n$}{number of rooms $j$}
    \printnomenclature
  \end{framed}
\end{table*}
\section{Introduction}\label{sec:introduction}
\acresetall

%%% motivation
Buildings consume one-third of global final energy \cite{IEA.2022} and produce \SI{27}{\percent} of total energy sector emissions. Almost half of the energy is used by \ac{HVAC} systems to heat or cool buildings \cite{Yang.2014}.
The energy consumption in buildings results from \ac{OB} and \ac{OTS} as they interact with the building's energy systems and require comfortable thermal conditions \cite{Hong.2017}. 
To reduce emissions, renewable energies can cover the energy demand of buildings \cite{VIEIRA2017308}.
As renewable energy sources are characterized by volatile energy generation, the buildings' electricity consumption could match this volatility.

%%% DR
Flexible electrical loads are pivotal for future energy systems in view of stability to cope with the increasing share of intermittent renewable energy sources like solar and wind energy. 
For exploiting flexible electric loads in buildings, the \ac{HVAC} operation can be integrated into \ac{DR} programs. 
\ac{DR} refers to the change of electricity demand in response to internal or external factors like the price of electricity \cite{4275494}. In the building sector, electrical \ac{HVAC} systems, like heat pumps or air conditioners, are suitable for \ac{DR}. They can exploit existing infrastructure like the building mass or hot water tanks to shift their electricity demand in time \cite{Dengiz2021_PHD}. Thus, they can significantly contribute to better utilization of renewable energy sources and simultaneously help to stabilize the electricity grid. In order to use \ac{HVAC} systems for \ac{DR}, optimized control strategies are necessary. 

%%% Occupant in multi-zone
In addition to \ac{DR}, designing the \ac{HVAC} operation tailored to the actual occupants' needs could significantly reduce energy use. For example, the average occupancy rates of offices are rarely over \SI{60}{\percent} \cite{Mahdavi.2008}. However, the \ac{HVAC} control in offices usually does not consider the actual occupancy of the individual rooms. This leads to unnecessary energy use in unoccupied periods. \SI{56}{\percent} of the energy consumed by buildings is used during unoccupied hours and \SI{44}{\percent} in occupied hours \cite{Masoso.2010}. To consider occupancy room-individually, multi-zone control strategies are required. 

%%% mpc
For the optimization of \ac{HVAC} to consider \ac{DR} and individual \ac{OTS}, advanced control strategies are required instead of standard thermostats \cite{Shaikh.2014}, for example, \ac{MPC} \cite{Drgona.2020} or heuristic control strategies \cite{Dengiz.2019}. \ac{MPC} finds the optimal input trajectory for the HVAC system's control outputs over a future time horizon by solving an optimization problem under consideration of system dynamics, forecasts, and constraints. Therefore, it requires a dynamic thermal building model and forecasts of \ac{OB} and weather \cite{Frahm.2022.2}. The development of models and forecasts can make \ac{MPC} less practicable and more expensive for real-world applications \cite{Drgona.2020}. 

%% heuristics
In contrast, heuristic control strategies are model- and forecast-free heuristic algorithms. They iteratively adjust the power consumption of \ac{HVAC} systems in order to archive certain goals. In order to do this, they use rule-based control mechanisms and heuristic algorithms that can adapt the \ac{HVAC} system's heat flows to internal and external signals. Their core advantage is that they do not require a building model to solve an optimization problem \cite{Dengiz.2019}. Thus, they apply to any building without significant adjustments.

\subsection{Related Work}
%%% intro and overview
A variety of different control strategies for \ac{HVAC} and evaluation methods are available in the literature. 
We compare the most relevant studies for the present paper in Tab.~\ref{tab:literature}, focusing on multi-zone control with heuristics algorithms and \ac{MPC}.

\begin{table*}[htb] 
    \centering
    \caption{Comparison of relevant papers studying approaches for demand response of HVAC systems}
    \begin{tabularx}{1\linewidth}{lXXXXXX}
        \toprule    
        Literature & Model-free control & Forecast-free control & Multi-zone control & Comparison with lower benchmark & Comparison with upper benchmark & Evaluation with different scenarios and parameters  \\ \midrule
        Dengiz et al., 2019 \cite{Dengiz.2019} & \cmark & \cmark & \xmark & \cmark & \cmark & \xmark \\ 
        Frahm et al., 2022 \cite{Frahm.2022.2} & \xmark & \xmark & \xmark & \cmark & \xmark & \xmark \\
        Zwickel et al., 2022 \cite{Zwickel.2022} & \xmark & \xmark & \cmark & \xmark & \cmark & \xmark \\ 
        Maddalena et al., 2022 \cite{Maddalena.2022} & \xmark & \xmark & \cmark & \cmark & \cmark & \cmark \\ 
        Hu et al., 2014 \cite{Hu.2014} & \xmark & \xmark & \cmark & \cmark & \cmark & \cmark \\ 
        Pedersen et al., 2018 \cite{Pedersen.2018} & \xmark & \xmark & \xmark & \cmark & \cmark & \xmark \\ 
        Blum et al., 2016 \cite{Blum.2016} & \xmark & \xmark & \cmark & \xmark & \cmark & \cmark \\ 
        Rodríguez et al., 2018 \cite{Rodriguez.2018} & \cmark & \cmark & \cmark & \cmark & \xmark & \xmark \\ 
        Nolting et al., 2019 \cite{Nolting.2019} & \cmark & \cmark & \xmark & \cmark & \xmark & \xmark \\ 
        Mork et al., 2022 \cite{Mork.2022} & \xmark & \xmark & \cmark & \cmark & \cmark & \xmark \\ 
        Michailidis et al., 2018 \cite{Michailidis.2018} & \cmark & \xmark & \cmark & \cmark & \xmark & \xmark \\ 
        Biyik et al., 2019 \cite{Biyik.2019} & \xmark & \xmark & \cmark & \cmark & \xmark & \xmark \\
        Peng et al. 2019 \cite{Peng.2018} & \cmark & \xmark & \cmark & \cmark & \xmark & \xmark \\
        Freund et al., 2021 \cite{Freund.2021} & \xmark & \xmark & \cmark & \cmark & \xmark & \cmark \\
        Korkas et al., 2016 \cite{Korkas.2016} & \cmark & \xmark & \cmark & \xmark & \cmark & \cmark\\
        \textbf{Present work} & \cmark & \cmark & \cmark & \cmark & \cmark & \cmark \\ 
        \bottomrule
    \end{tabularx}
    \label{tab:literature}
\end{table*}

In all evaluated studies, the objective is to reduce energy costs while satisfying \ac{OTS}. As a result, most authors compare the energy costs and thermal comfort as \ac{KPIs} for the controllers. However, the performance results not only from the controller itself but also from the evaluation environment. 
Overall, the performance of controllers depends on three major variables: 
\begin{itemize}
    \item \textbf{controller}: the logic of the controller and the information it processes in its decision-making process,
    \item \textbf{data}: the data and scenarios used for evaluation, including weather, price, and occupancy data,
    \item \textbf{parameters}: the model parameters of the evaluation environment, such as thermal capacitance.
\end{itemize}

The three most significant differences between controllers are (i) whether they require a model, (ii) forecasts, and (iii) if they can control multiple zones. Regarding the controller evaluation, we investigate if upper and lower benchmarks as well as different scenarios and different parameters were used for evaluation.

%%% Model-based
Most studies in the literature use model-based approaches, such as \ac{MPC}, as they can find the optimal solution of an optimization problem \cite{Dengiz2021_PHD}. 
\ac{MPC} has gained significant importance for building control in the context of \ac{DR}.
Most authors use \ac{MPC} for controlling \ac{HVAC} systems, e.g. Maddalena et al. \cite{Maddalena.2022},  Hu et al. \cite{Hu.2014}, Pedersen et al. \cite{Pedersen.2018}, Blum et al. \cite{Blum.2016}, Mork et al. \cite{Mork.2022}, Frahm et al. \cite{Frahm.2022.2}, Zwickel et al. \cite{Zwickel.2022}, Biyik et al. \cite{Biyik.2019}, and Freund et al. \cite{Freund.2021}. While model-based approaches generally yield adequate results, they can suffer from execution times and require modeling the thermal behavior of a building, which is the most complex task \cite{Drgona.2020}. 

%%% Model-free
Fewer studies use model-free control strategies. Compared to model-based strategies, the controller design process is significantly simplified, as no building-specific model is required. Model-free control algorithms can be found in the studies of Dengiz et al. \cite{Dengiz.2019}, Rodriguez et al. \cite{Rodriguez.2018}, Nolting et al. \cite{Nolting.2019}, and Michailidis et al. \cite{Michailidis.2018}, Peng et al. \cite{Peng.2018}, and Korkas et al. \cite{Korkas.2016}. These approaches are rule-based control mechanisms that, in a few cases, are also combined with a heuristic approach for optimizing an objective function. 

%%% Forecast-free
Another essential requirement for most of the optimized control approaches is the availability of forecasts. Some of the model-free approaches do not rely on any forecast, such as Dengiz et. al. \cite{Dengiz.2019}, Rodriguez et al. \cite{Rodriguez.2018}, or Notling et al. \cite{Nolting.2019}.

%%% single zone vs. multi-zone and multiple buildings
Our literature review emphasizes the use of control algorithms for multiple zones (see Tab.~\ref{tab:literature}).
There are also control approaches in the literature that consider only buildings with one thermal zone. 
However, this would assume a uniform temperature in the whole building. 
As a result, the consideration of multiple zones is closer to the real thermal behavior of buildings.
On the one hand, multi-zone control increases the complexity of the control problem. 
On the other hand, a controller that is designed for multiple rooms can consider room individual \ac{OTS} and \ac{OB}. In addition, multi-zone control strategies for multiple rooms can often be extended to multiple buildings. For example, Korkas et al. \cite{Korkas.2016} demonstrate the control of three different buildings, where each building has ten zones.

% Evaluation
When evaluating the performance of the developed control approach, most studies use a conventional baseline control approach, like simple rule-based control, hysteresis-based two-point controller, or a \ac{PI} controller as a lower benchmark. 
Rarely, a heuristic controller is compared to a lower benchmark (baseline controller) and also an upper benchmark (optimal-based controller). 
Such an elaborated evaluation can be found in Hu et al. \cite{Hu.2014} who compare a heuristic, baseline, and \ac{MPC} controller, and also two variants of each. 

To find out whether control strategies are also suitable for various conditions, it is required to compare different scenarios, data sets, and parameters in the evaluation environment. Over several months, Freund et al. \cite{Freund.2021} compared three different building sections, with different controllers and up to seven rooms in each section. Korkas et al. \cite{Korkas.2016} evaluated \ac{RBC} and an optimization-based strategy in three different buildings with individual occupancy schedules. Hu et al. \cite{Hu.2014} demonstrated the impact of different thermal parameters and varying window operation schedules on the control performance. Changes in weather conditions and their impact on the controllers (\ac{MPC} and \ac{PI}) can be found in Maddalena et al. \cite{Maddalena.2022}. Blum et al. \cite{Blum.2016} showed the impact of the thermal mass on \ac{DR} provision with optimization-based strategies. 

In summary, the literature review shows a demand for multi-zone control to consider occupant-oriented \ac{DR} room-individually. To address occupant-oriented \ac{DR}, most authors apply an optimization-based approach, such as \ac{MPC}. However, \ac{MPC} requires a thermal building model and it is a complex task finding sufficient models for various buildings. Fewer studies use model-free strategies with heuristic approaches. In addition, we obtained a research gap for an elaborative evaluation of multi-zone heuristic control strategies with different scenarios and parameters as well as upper and lower benchmarks.

\subsection{Contribution of this Paper}
%\textit{How can we control a building to provide \ac{DR} while considering room-individual \ac{OTS}?}

The two main contributions of the present paper are (i) the introduction of a novel heuristic multi-zone control approach, called \ac{PSC}, and (ii) the evaluation in a versatile evaluation environment with different scenarios, parameters, and data sets. 

\ac{PSC} combines external factors (e.g. electricity price) and internal factors (temperatures of different zones in the building) to determine when and how much electricity should be consumed for the generation of heat flows. The approach is model-free and does not need any forecasts. To the best of our knowledge, our study is the only one that introduces a novel control approach for buildings with multiple zones that does not need any model or forecasts and that allows for a coordinated coupling of multiple buildings. 
This is because of its capability to use any external factor for deriving the \ac{HVAC} control output. 
Our study is the first that evaluates an introduced model-free and forecast-free control algorithm by using a lower and upper benchmark that is evaluated with different scenarios and parameters (see Tab.~\ref{tab:literature}). 

To evaluate the \ac{PSC} control performance in terms of \ac{OTS} and energy costs, we compare four different control strategies in a multi-zone thermal building model. 
In the evaluation, we use two different parameter sets for high and low capacitance and two scenarios with different degrees of variable room usage. In the base scenario, the temperature range is scheduled between comfort and standby mode. The second scenario also allows room-individual temperature ranges, based on the use case for each room. For comparison, we use an ideal, error-free \ac{MPC}, a simplified version of \ac{PSC} without storage factor, called \ac{PC} that is based on an approach from the literature \cite{Dengiz.2019}, and a hysteresis-based two-point controller as upper and lower benchmarks. 
We publish the four developed controllers and the evaluation environment as commented Python code in an open-source repository. 
This publication enables researchers to (i) directly apply our developed control strategies in other scenarios and environments, (ii) test \ac{OB} models regarding control metrics \cite{Dong.2022}, and (iii) develop new control strategies on the room-individual level. 

\subsection{Structure of this Paper}
We develop and implement four different control strategies and an evaluation environment in the present work. We present the models in Sec.~\ref{sec:model}, the controllers in Sec.~\ref{sec:control}, the evaluation environment in Sec.~\ref{sec:evaluation}, and results and discussion in Sec.~\ref{sec:results}. Finally, we conclude the paper in Sec.~\ref{sec:conclusion}.

\section{Models}\label{sec:model}
The modeling section Sec.~\ref{sec:model} is separated into three parts: the thermal building model in Sec.~\ref{sec:building_model}, the heat pump in Sec.~\ref{sec:heat pump model}, and \ac{OTS} in Sec.~\ref{sec:OTS}.
We use the models for the evaluation environment in Sec.~\ref{sec:evaluation} and for the internal model of \ac{MPC} (see Sec.~\ref{sec:MPC}).

\subsection{Multi-Zone Thermal Building Model}\label{sec:building_model}
In this section, we develop a multi-zone thermal building model to evaluate room-individual control strategies.
The model applies the \ac{RC} analogy to describe the heat flows between temperature nodes by resistors $R$ and thermal dynamics by capacitors $C$.
First, we derive the \ac{RC} analogy and then apply it to the multi-zone model for evaluation and \ac{MPC}.

\subsubsection*{Derivation of the RC analogy}

In general, the \ac{RC} analogy is used to model thermal behavior in buildings by simplifying the laws of thermodynamics and heat transfer \cite{Frahm.2022.1}. In analogy to electric networks, the thermal behavior of a building is determined by resistors and capacitors.

Thermal resistors $R$ describe possible paths of heat flows and their rate of transmitted energy. As illustrated in Eq.~\eqref{eq:resistor}, the temperature $T$ is analogous to the voltage $U$ and the heat flow $\dot{Q}$ to the electric current $I$.

\begin{equation}
\begin{aligned}
    \Delta U &= R \cdot I \\
    \Rightarrow \Delta T &= R \cdot \dot{Q}
\end{aligned}
\label{eq:resistor}
\end{equation}

As shown in Eq.~\eqref{eq:capacitor}, thermal capacitors $C$ quantify the thermal capacity of thermal elements. In a building, these elements can be for example the air, walls, or furniture. 

\begin{equation}
\begin{aligned}
    C \dv{U(t)}{t} &= I_\mathrm{in} (t) - I_\mathrm{out} (t) \\
    \Rightarrow  C \dv{T(t)}{t} &= \dot{Q}_\mathrm{in} (t) - \dot{Q}_\mathrm{out} (t)
\end{aligned}
\label{eq:capacitor}
\end{equation}

Analog to a voltage $U$, the temperature $T$ of the capacitor $C$ describes the thermal state of the thermal element. 
To capture the thermal state of an entire building with a model, the building is simplified into a discrete number of states. This approach is also called the lumped capacitance method.
The number of states and the model structure needs to be adequately defined so that the thermal states can sufficiently describe the heat transfer effects associated with thermal elements in contact \cite{Koeln.2018}.

\subsubsection*{Multi-zone model for evaluation and \ac{MPC}}

\begin{figure}[htb]
\centering
    \def\svgwidth{1.0\linewidth}  
    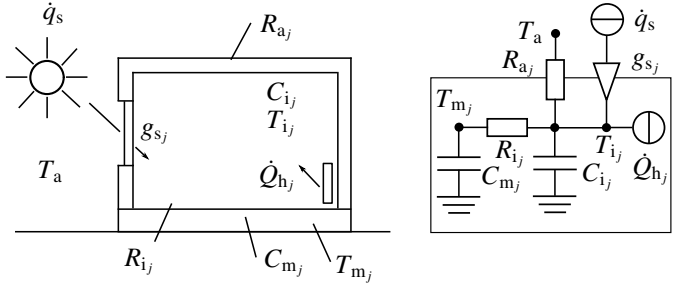
    \caption{Thermal building model for each room  $j$}
    \label{fig:basicModel}
\end{figure}

We illustrate our thermal building model structure, obtained from the literature \cite{Madsen.1995,Frahm.2022.4}, in Fig.~\ref{fig:basicModel}. In this model, we use two states for each room $j$\footnote{We use the subscript $j$ as a room index for $n$ rooms: $j$ ($j = 1 \dots n$).}, the inside air temperature $T_{\mathrm{i}_j}$ and the thermal mass $T_{\mathrm{m}_j}$. 
These two states are connected with resistors $R$, capacitors $C$, and heat flows $\dot{Q}$.

The two temperature nodes $T_{\mathrm{i}_j}$ and $T_{\mathrm{m}_j}$ are connected by the inside resistor $R_{\mathrm{i}_j}$. Furthermore, the inside air temperature node $T_{\mathrm{i}_j}$ is connected to the outside air temperature $T_\mathrm{a}$ using the outside resistor $R_{\mathrm{a}_j}$. The inside air temperature $T_{\mathrm{i}_j}$ is also effected by heat flows from the sun $\dot{Q}_{\mathrm{s}_j}$ and the heating $\dot{Q}_{\mathrm{h}_j}$. The solar heat flow $\dot{Q}_{\mathrm{s}_j}$ results from the global radiation $\dot{q}_\mathrm{s}$ and the solar heat gain coefficient $g_{\mathrm{s}_j}$.

Based on Fig~\ref{fig:basicModel}, each room is mathematically defined by the two differential equations Eq.~\eqref{eq:dT_i} and \eqref{eq:dT_m}. Applying this structure in Fig.~\ref{fig:basicModel} to each room $j$ ($j = 1 \dots n$) results in a multi-zone model \cite{Frahm.2022.4} (see Fig.~\ref{fig:floorplan} in Sec.~\ref{sec:evaluation}).
\begin{align}
    C_{\mathrm{i}_j} \dv{T_{\mathrm{i}_j}}{t} &= 
    \frac{T_{\mathrm{m}_j} - T_{\mathrm{i}_j}}{R_{\mathrm{i}_j}} + 
    \frac{T_{\mathrm{a}} - T_{\mathrm{i}_j}}{R_{\mathrm{a}_j}} + g_{\mathrm{s}_j} \dot{q}_\mathrm{s} + \dot{Q}_{\mathrm{h}_j} \label{eq:dT_i} \\ 
    C_{\mathrm{m}_j} \dv{T_{\mathrm{m}_j}}{t} &= \frac{T_{\mathrm{i}_j} - T_{\mathrm{m}_j}}{R_{\mathrm{i}_j}} \label{eq:dT_m}
\end{align}

Each room needs to be heated by a heat flow $\dot{Q}_{\mathrm{h}_j}$, provided by the heat pump. 

\begin{equation}
    \sum_{j=1}^{n} \dot{Q}_{\mathrm{h}_j} = \dot{Q}_{\text{h}} 
\end{equation}

The heat pump generates the sum of heat flows $\dot{Q}_{\text{h}}$ from electric power $P_\text{el}$, as described in the following.

\subsection{Heat Pump Model}\label{sec:heat pump model}

The heat pump transforms electrical power $P_\text{el}$ to a heat flow $\dot{Q}_{\text{h}}$ with a coefficient of performance $\varepsilon_\text{h}$. 

\begin{equation}
    P_\text{el} = \frac{\abs{\dot{Q}_{\text{h}}}}{\varepsilon_\text{h} } \label{eq:heatpumpheatflowsrelation}
\end{equation}

In this study, a air-sourced heat pump is considered where the coefficient of performance depends on the temperature of the supplied air (the ambient temperature) $T_\text{a}$: $\varepsilon_\text{h} = \varepsilon_\text{h} (T_\text{a})$. %Also, the maximum electrical power $P_\text{max}$ depends on the ambient temperature.

The heat pump can modulate its power consumption $P_\text{el}$ with $\chi_\text{mod}$ between \SI{20}{\percent} and \SI{100}{\percent} (and thus the resulting heat flow $\dot{Q}_{\text{h}}$). 

\begin{align}
    P_\text{el} = \chi_\text{mod} \cdot P_\text{max} \label{eq:heatpumpupower} \\
    \chi_\text{mod} \in \{0,[0.2,1]\} \label{eq:heatpumputilization}
\end{align}

In our model, the entire electrical power $ P_\text{el}$ needs to be bought from the grid. 
\begin{equation}
    P_\text{buy} =  P_\text{el}
\end{equation}

\subsection{\acf{OTS} Model}\label{sec:OTS}
In this section, we define the temperature ranges $[T_{\mathrm{lb}_j}(t), T_{\mathrm{ub}_j}(t)]$ based on international standards for \acf{OTS} modeling. The three most frequently cited \ac{OTS} standards are \textit{ASHRAE Standard 55} \cite{ASHRAE55}, \textit{ISO 7730:2005} \cite{ISO7730}, and \textit{EN 16798-1:2019} \cite{EN16798}. These standards are fundamentally based on the \ac{PMV} standard scale, which was first introduced by Fanger's model \cite{Fanger.1970}. 

The \ac{PMV} is a static model evaluated from a large group of people with a given combination of thermal environmental and personal parameters. These parameters include metabolic activity, clothing, air temperature, radiant temperature, air velocity, and relative humidity. In a survey, occupants express their thermal sensations on a scale from -3 (too cold) to +3 (too warm), where 0 is optimum. 
Fanger also developed an equation that relates the \ac{PMV} to the \ac{PPD}.

The \ac{OTS} level can be selected between different \ac{PMV} boundaries, e.g. $\pm 0.2$ for level I or $\pm 0.7$ for level III. The standard \ac{OTS} guidelines aim for a \ac{PMV} from -0.5 to +0.5 (\ac{OTS} level II, see Tab.~\ref{tab:comfort}). Wider temperature limits result in lower energy consumption of \ac{HVAC} systems, while smaller limits reduce the \ac{PPD}. 

\begin{table}[htb]
    \centering
    \caption{\ac{OTS} categories, obtained from \href{https://comfort.cbe.berkeley.edu/EN}{CBE Thermal Comfort Tool} \cite{Tartarini.2020} with EN-16798 and winter clothings}
    \begin{tabularx}{1\linewidth}{XXXXX}
        \toprule
        \ac{OTS} & PMV & PPD & $T_\mathrm{lb}$ & $T_\mathrm{ub}$ \\
        \midrule
        I & $\pm 0.2$ & $<\SI{6}{\percent}$ & 22.6 \si{\celsius} & 24.0 \si{\celsius}  \\
        II & $\pm 0.5$ & $<\SI{10}{\percent}$ & 21.5 \si{\celsius} & 25.0 \si{\celsius} \\
        III & $\pm 0.7$  &  $<\SI{15}{\percent}$ & 20.7 \si{\celsius} & 25.8 \si{\celsius} \\
        off & - & - & 16.0 \si{\celsius} & 30.0 \si{\celsius} \\
        \bottomrule 
    \end{tabularx}
    \label{tab:comfort}
\end{table}

Based on these \ac{OTS} levels in Tab.~\ref{tab:comfort}, we calculate the corresponding lower $T_\mathrm{lb}$ and upper $T_\mathrm{ub}$ temperature limits that are required for the scenarios in Sec.~\ref{sec:Scenarios}.
For the calculation of the temperature limits, we use the \href{https://comfort.cbe.berkeley.edu/EN}{CBE Thermal Comfort Tool} \cite{Tartarini.2020} with EN-16798 standard and winter clothing. 
In this tool, we set the mean radiant temperature equal to the air temperature $T_\text{i}$. This implies the assumption that the operative temperature is close to the air temperature.
For more information about the operative temperature, we refer to our previous work \cite{Frahm.2022.1}.
The resulting temperature limits for different levels of \ac{OTS} are presented in Tab.~\ref{tab:comfort}.

\section{Control Strategies}\label{sec:control}

This section describes the development of four different control strategies:
\ac{MPC} in Sec.~\ref{sec:MPC}, \ac{PSC} and \ac{PC} in Sec.~\ref{sec:PSC}, and hysteresis-based two point control in Sec.~\ref{sec:Heuristic}. In general, the four control strategies apply to heating or  cooling\footnote{For both cases, we use the generic term \textit{heat flows}. A heat flow is the rate of net heat energy transfer between hot and cold sides and can be positive or negative for heating or cooling, respectively.}. 

The objective is to minimize the electricity costs given by a time-variable electricity price while ensuring the \ac{OTS}. We develop the \ac{PSC} as a novel control methodology for occupant-oriented demand-response with multi-zone building control. The \ac{MPC} and hysteresis-based two-point controller are used as upper and lower benchmarks, respectively. 

For the implementation of all controllers, we used Python and the do-mpc toolbox \cite{dompc.2017}.
The \ac{MPC} has a prediction horizon of 16 hours and is solved with CasADi \cite{Casadi.2019} in Python.

%%%%%%%%%%%%%%%%
%%%%%%%%%%%%%%%%
%%%%  MPC %%%%%%
%%%%%%%%%%%%%%%%
%%%%%%%%%%%%%%%%
\subsection{\acf{MPC}}\label{sec:MPC}

In contrast to the other control strategies, \ac{MPC} requires a model. The model is obtained from Sec.\ref{sec:model} and then reformulated to the general control notation in Eq.~\eqref{eq:model_cont}, 
\begin{subequations}
\begin{align}
    \dv{x(t)}{t} &= f \left(  x(t), u(t),  p_\text{tv}(t)  \right) \\
    0 &= g \left(  x(t), u(t),  p_\text{tv}(t)  \right) \\
    y(t) &= h \left(  x(t), u(t),  p_\text{tv}(t)  \right)
\end{align}
\label{eq:model_cont}
\end{subequations}
where $x$ are states, $u$ control inputs, $p_\text{tv}$ time-variable paramters, and $y$ control outputs.

For the implementation of the model on a computer or a micro-controller using the do-mpc Python toolbox \cite{dompc.2017}, the time-continuous formulation from Eq.~\eqref{eq:model_cont} is discretized to Eq.~\eqref{eq:model_disc}:
\begin{subequations}
\begin{align}
    x[k+1] &= f \left(  x[k], u[k],  p_{\text{tv}}[k]  \right) \\
    0 &= g \left(  x[k], u[k],  p_{\text{tv}}[k]  \right) \\
    y[k] &= h \left(  x[k], u[k],  p_{\text{tv}}[k]  \right)
\end{align}
\label{eq:model_disc}
\end{subequations}

The states, control inputs, time-variable parameters, and control outputs are obtained from the model in Sec.\ref{sec:model} with:

\begin{align}
    x[k] &= \left(  T_{\text{i}_1}[k] \; T_{\text{m}_1}[k] \; \dots \; T_{\text{i}_n}[k] \; T_{\text{m}_n}[k] \right)^\mathsf{T} \\
    y[k] &= \left(  T_{\text{i}_1}[k] \; \dots \; T_{\text{i}_n}[k] \right)^\mathsf{T} \\
    u[k] &= \left( \dot{Q}_{\text{h}_1}[k] \; \dots \; \dot{Q}_{\text{h}_n}[k] \right)^\mathsf{T} \\
    p_{\text{tv}}[k] &= \left(  T_\text{a}[k] \; \dot{q}_\text{s}[k] \; p_\text{buy}[k] \right)^\mathsf{T} \label{eq:p_tv}
\end{align}

\begin{itemize}
    \item $x(t)$, $x[k]$: states,
    \item $y(t)$, $y[k]$: control outputs (measurements),
    \item $u(t)$, $u[k]$: control inputs,
    \item $p_\text{tv}(t)$, $p_{\text{tv}}[k]$: time-variable parameters.
\end{itemize}

The MPC uses the reformulated, discretized model from Eq.~\eqref{eq:model_disc}~-~\eqref{eq:p_tv} and optimizes a cost function $C$ in Eq.~\eqref{eq:cost_function},

\begin{equation}
    C = 
    \underbrace{
    m(x[N+1]) 
    }_\text{Meyer term}
    + \sum_{k=0}^{N} \left( 
    \underbrace{
    l(x[k], u[k],  p_{\text{tv}}[k]) 
    }_\text{Lagrange term}
    + 
    \underbrace{
    \Delta u[k]^\mathsf{T} R \Delta u[k]
    }_\text{R-term}
    \right)
    \label{eq:cost_function}
\end{equation}
where the Meyer term $m(\cdot)$ defines costs of the terminal state $x[N+1]$,  the Lagrange term $l(\cdot)$ the costs of each stage $k$, and the R-term costs for changes in inputs.

Finally, the overall MPC optimization is formulated in Eq.~\eqref{eq:MPC}, including a cost function in Eq.~\eqref{eq:opti:costs}, the model equations in Eq.~\eqref{eq:opti:x0}~-~\eqref{eq:opti:stat}, and constraints in Eq.~\eqref{eq:opti:x}~-~\eqref{eq:opti:u}.

\begin{subequations}
    \begin{align}
    \min_{ \mathbf{x}[0:N+1], \mathbf{u}[0:N] } &
    \sum_{k=0}^{N} \left( l(x[k], u[k],  p_{\text{tv}}[k]) + \Delta u[k]^\mathsf{T} R \Delta u[k] \right) \label{eq:opti:costs} \\
    &\textrm{subject to} \; \forall k \in [0,N]: \notag \\
    &x_0 = \hat{x}_0  & \label{eq:opti:x0} \\
    &x[k+1] = f \left(  x[k], u[k],  p_{\text{tv}}[k]  \right) \label{eq:opti:dyn}  \\
    & 0 =g \left(  x[k], u[k],  p_{\text{tv}}[k]  \right)  \\
    & y[k] = h \left(  x[k], u[k],  p_{\text{tv}}[k]  \right) \label{eq:opti:stat} \\
    & y_\mathrm{lb}[k] \leq y[k] \leq y_\mathrm{ub}[k] \label{eq:opti:x}  \\
    & u_\mathrm{lb}[k] \leq u[k] \leq u_\mathrm{ub}[k] \label{eq:opti:u}
    \end{align}
\label{eq:MPC}
\end{subequations}

Under the consideration of constraints, the main purpose of the optimization is to reduce energy costs that are formulated in the Lagrange term $l(\cdot)$ in Eq.~\eqref{eq:lagrange},  

\begin{equation}
    l(x[k], u[k],  p_{\text{tv}}[k]) = P_\text{buy}[k] \cdot p_\text{buy}[k] \label{eq:lagrange}
\end{equation}
where $p_\text{buy}[k]$ is the dynamic energy price at step $k$ and $P_\text{buy}[k]$ the bought power. This power depends on the sum of heat flows from the control inputs $u$ and the coefficient of performance (see Sec~\ref{sec:heat pump model}).

In our control problem, we require no specific terminal-state $x[N+1]$, which eliminates the Meyer term in Eq.~\eqref{eq:opti:costs}. Instead, we need our outputs $y$, a subset of the states $x$, to remain within lower and upper bounds $y_\mathrm{lb}$ and $y_\mathrm{ub}$, as formulated in Eq.~\eqref{eq:opti:x}. 
In Eq.~\eqref{eq:opti:u}, the control inputs $u$ are constrained to minimal and maximal values $u_\mathrm{lb}$ and $u_\mathrm{ub}$ to ensure physical feasibility. 

%%%%%%%%%%%%%%%%
%%%%%%%%%%%%%%%%
%%%%  PSC %%%%%%
%%%%%%%%%%%%%%%%
%%%%%%%%%%%%%%%%

\subsection{\acf{PSC}}\label{sec:PSC}

The \ac{PSC} is a heuristic control algorithm for modulating \ac{HVAC} or heat pump heat flows $\dot{Q}_{\mathrm{h}_j}$ in a multi-zone building without considering a building model. It essentially consists of 4 steps which it executes in every time slot $k$.
\begin{enumerate}
  \item Determine the price factor $\chi_\text{p} [k]$ (similar to \cite{Dengiz.2019}).
  \item Determine the storage factor $\chi_\text{s} [k]$.
  \item Calculate the modulation degree $\chi_\text{mod}$ using the price factor $\chi_\text{p}[k]$ and the storage factor $\chi_\text{s}[k]$.
   \item Distribute the generated heat flow to the different rooms of the multi-zone building.
\end{enumerate}

\subsubsection*{Price Factor}
\begin{figure*}[htb]
\centering
    \def\svgwidth{\linewidth}  
    %% Creator: Inkscape 1.2.1 (9c6d41e410, 2022-07-14), www.inkscape.org
%% Accompanies image file 'EDF.pdf' (pdf, eps, ps)
\begingroup%
  \makeatletter%
  \providecommand\color[2][]{%
    \errmessage{(Inkscape) Color is used for the text in Inkscape, but the package 'color.sty' is not loaded}%
    \renewcommand\color[2][]{}%
  }%
  \providecommand\transparent[1]{%
    \errmessage{(Inkscape) Transparency is used (non-zero) for the text in Inkscape, but the package 'transparent.sty' is not loaded}%
    \renewcommand\transparent[1]{}%
  }%
  \providecommand\rotatebox[2]{#2}%
  \newcommand*\fsize{\dimexpr\f@size pt\relax}%
  \newcommand*\lineheight[1]{\fontsize{\fsize}{#1\fsize}\selectfont}%
  \ifx\svgwidth\undefined%
    \setlength{\unitlength}{563.48591482bp}%
    \ifx\svgscale\undefined%
      \relax%
    \else%
      \setlength{\unitlength}{\unitlength * \real{\svgscale}}%
    \fi%
  \else%
    \setlength{\unitlength}{\svgwidth}%
  \fi%
  \global\let\svgwidth\undefined%
  \global\let\svgscale\undefined%
  \makeatother%
  \begin{picture}(1,0.30189941)%
    \lineheight{1}%
    \setlength\tabcolsep{0pt}%
    \put(0,0){\includegraphics[width=\unitlength,page=1]{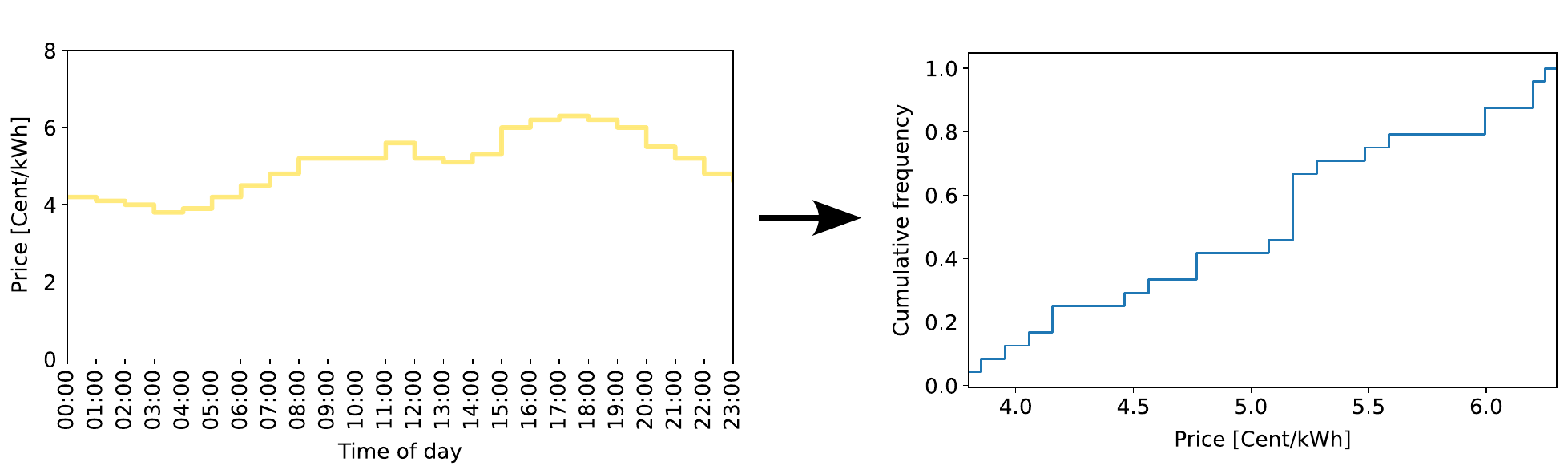}}%
    \put(0.23297937,0.28960846){\color[rgb]{0,0,0}\makebox(0,0)[lt]{\lineheight{1.25}\smash{\begin{tabular}[t]{l}$p(t)$\end{tabular}}}}%
    \put(0.78686908,0.28810538){\color[rgb]{0,0,0}\makebox(0,0)[lt]{\lineheight{1.25}\smash{\begin{tabular}[t]{l}$\widehat{F}(p)$\end{tabular}}}}%
  \end{picture}%
\endgroup%

    \caption{Empirical distribution function of the electricity prices}
    \label{fig:EDF}
\end{figure*}

The price factor quantifies the heating tendency of the heat pump based on the current price $p_\text{buy}[k]$. To obtain the price factor $\chi_\text{p}$, the algorithm calculates the empirical distribution function $\widehat{F}(p_\text{buy}[k])$ for the future electricity prices $p_\text{buy}[k]$ of the next 24 hours at the beginning of each day. We assume that we have an electricity tariff with predetermined prices for the next 24 hours (for more information see Sec.~\ref{sec:data}). At every time slot $k$ of the day, the value of the $\widehat{F}(p_\text{buy}[k])$ is calculated for the current price $p_\text{buy}[k]$. 
The calculation of the empirical distribution function $\widehat{F}(p)$ is illustrated in Fig.~\ref{fig:EDF} exemplarily for one day.
$\widehat{F}(p_\text{buy}[k])$ quantifies the share of electricity prices for the current day that has a lower or equal value compared to the price $p_\text{buy}[k]$ of the current time slot $k$. 
\ac{PSC} sets the price factor at time slot $k$ as in Eq.~\eqref{eq:price_factor}. A low price results in a high price factor (because of a low value of $\widehat{F}(p_\text{buy}[k])$) and vice versa. 
\begin{equation}
\chi_\text{p}[k]= 1 - \widehat{F}(p_\text{buy}[k])
\label{eq:price_factor}
\end{equation}

\subsubsection*{Storage Factor}

For the calculation of the storage factor $\chi_\text{s} [k]$, the state of thermal charge $S_{j}[k]$ from Eq.~\eqref{eq:Storage_percentage} is needed for each room.
The state of thermal charge $S_{j}[k]$ quantifies the stored thermal energy room individually, based on the temperature of room $j$ of the last time slot $T_{\text{i}_j}[k-1]$ \footnote{Although the \ac{PSC} method is applicable for heating or cooling heat flows, we explain this method exemplarily for the heating case in the following.}. 

\begin{equation}
  S_j[k] =  \frac{ T_{\text{i}_j}[k-1] - T_{\text{lb}_j}[k]   }{T_{\text{ub}_j}[k] -T_{\text{lb}_j}[k]} 
  \label{eq:Storage_percentage_theo}
\end{equation}

To remain the value of $S_{j}[k]$ between 0 and 1, we cap values of $S_j[k]$ larger than 1 and smaller than 0 with $S_j^*[k]$. 

\begin{equation} 
\label{eq:Storage_percentage}
S_j^*[k]= 
\begin{cases}
S_j[k] & \text{for} \quad 0 \leq S_j[k] \leq 1  \\
1 & \text{for} \quad S_j[k] > 1 \\
0 & \text{for} \quad S_j[k] < 0 \\
\end{cases}
\end{equation}

The lower bound of the comfort temperature $T_{\text{lb}_j}[k]$ is subtracted from the temperature of room $j$ of the last time slot $T_{\text{i}_j}[k-1]$. This is then divided by the distance between the upper bound of the comfort temperature $T_{\text{ub}_j}[k]$ and the lower bound $T_{\text{lb}_j}[k]$ 

$S_j^*[k]=1$ means that the thermal energy storage of this room is sufficiently full and there is no necessity for applying heat flows to the room \footnote{As we are considering heating in the present work it has to be noted that full thermal storage, in this case, means, that the temperature in the room is high enough. For cooling, in contrast, a low enough temperature would mean full storage.}. 
If the temperature of the room $T_{\text{i}_j}[k-1]$ is lower than the lower limit comfort temperature $T_{\text{lb}_j}$, the state of thermal charge $S_{j}^*[k]$ is set to 0. In the heating case, this results in empty thermal storage as the temperature in the room is too low.

After having determined the state of thermal charge $S_{j}^*[k]$ for every room $j$ ($j=1 \dots n$), the algorithm calculates the storage factor $\chi_\text{s} [k]$ by using Eq. (\ref{eq:Storage_factor}). If the temperatures in the different rooms are close to the upper limit, their corresponding state of thermal charge will be high resulting in a low storage factor $\chi_\text{s} [k]$ and vice versa.  
\begin{equation} 
\label{eq:Storage_factor}
\chi_\text{s} [k] = 1 - \frac{\sum_{j=1}^{n} S_{j}^*[k]}{n}
\end{equation}

\subsubsection*{Modulation Degree of the HVAC system}
The third step of the algorithm is the calculation of the heat pump's modulation degree and thus the heat flow and the electrical power (see Eq.~\eqref{eq:heatpumpupower}) using Eq.~\eqref{eq:modDegree}. The modulation degree $\chi_\text{mod}[k]$ results from the multiplication of the price factor $\chi_\text{p}$ and storage factor $\chi_\text{s}$. Because both factors can have values between 0 and 1, the modulation degree $\chi_\text{mod}[k]$ likewise varies between 0 and 1. We choose a multiplication of the two factors instead of a weighted sum as this leads to better results in our case studies. Based on the modulation degree, Eq.~\eqref{eq:heatpumpheatflowsrelation} and Eq.~\eqref{eq:heatpumpupower} calculate the generated heat flows and electrical power.
\begin{equation} 
\label{eq:modDegree}
\chi_\text{mod} [k] = \chi_\text{p} [k] \cdot \chi_\text{s} [k]
\end{equation}

Two factors influence the heat pump's power output. A high electricity price leads to a low price factor which leads to low values of the modulation degree. This results in low electricity consumption at that time. On the contrary, a low price leads to a high price factor which incentives the heat pump to heat the room. This is desired as we want to generate heat flows when the electricity prices are low. 

Next to the price factor, the storage factor impacts the generated heat flows and thus consumed electricity. If the temperatures in the rooms are generally high, the storage factor has low values because of the high values of the state of thermal charge $S_{j}^*[k]$. A low storage factor leads to low power consumption and vice versa. This is also a desired property of the control algorithm. If the room temperatures are already high, there is no urgent need for heating whereas low room temperatures tend to lead to higher generation of heat flow using the \ac{PSC} algorithm. 

The described approach of \ac{PSC} with the multiplication of the two factors is just the basic version of the algorithm. \ac{PSC} can easily be adjusted and enhanced by adding additional rule-based control actions that may depend on the used comfort model. The core of the algorithm, however, is always the use of the price and storage factor for determining the heat pump's modulation degree.

\subsubsection*{Distribution of Heat Flows}
In the final step, the algorithm distributes the generated heat flows to the different rooms $j$ ($j=1 \dots n$). To do this, the caused thermal discomfort of each room $d_{\text{c}_j}[k]$ due to possibly too low temperatures is determined. If the temperature of a room from the previous time slot $T_{\text{i}_j}[k-1]$ is lower than the lower temperature limit $T_{\mathrm{lb}_j}[k]$, Eq.~\eqref{eq:discomfort} and Eq.~\eqref{eq:discomfortAll} quantify the caused discomfort of the room $j$ and the total caused discomfort $d_\text{c,total}[k]$ from Eq.~\eqref{eq:total_dc}.

\begin{align} 
\label{eq:discomfort}
d_{\text{c}_j}[k] &= T_{\text{i}_j}[k-1] - T_{\mathrm{lb}_j}[k] \\
d_\text{c,total}[k] &=\sum_{j=1}^n d_{\text{c}_j}[k] 
\label{eq:total_dc}
\end{align}

Based on the total caused discomfort $d_\text{c,total}[k]$, the \ac{PSC} algorithm distributes the generated heat flows $\dot{Q}_{h}$ at each next step $k$ to each room $j$ with $\dot{Q}_{\mathrm{h}_j}$ using Eq. (\ref{eq:discomfortAll}). This mechanism assures that especially rooms that have low temperatures get larger heat flows than rooms with less need for heating.
If the heat pump generates heat flows although no room has violated its temperature boundaries in the last time slot $k$, it equally distributes the generated heat flows to every room.   

\begin{equation} 
\label{eq:discomfortAll}
\dot{Q}_{\mathrm{h}_j}[k] =  \frac{d_{\text{c}_j}[k]}{\sum_{j=1}^n d_{\text{c}_j}[k]} \cdot \dot{Q}_{h}[k]
\end{equation}

Overall, \ac{PSC} executes the four mentioned steps for every time slot of the day while updating the empirical distribution function of the prices at the beginning of each day.

%%%%%%%%%%%%%%%%%
%%%%%%%%%%%%%%%%%
%%%%  Hyst %%%%%%
%%%%%%%%%%%%%%%%%
%%%%%%%%%%%%%%%%%

\subsection{Hysteresis-based Two-point Controller
}\label{sec:Heuristic}
The hysteresis-based two-point control serves as the lower benchmark for the evaluation. This is a conventional control strategy for heating or cooling devices that heats a room until the upper temperature limit $T_{\mathrm{ub}_j}[k]$ is reached. Afterward, the device switches off and waits until the temperature in the room reaches the lower limit $T_{\mathrm{ub}_j}[k]$. This triggers the control system to start heating again. We use an adaptive hysteresis that uses the upper and lower temperature limits $[T_{\mathrm{lb}_j}(t), T_{\mathrm{ub}_j}(t)$ depending on the scenarios. These predefined temperature limits for \ac{OTS} are described in the evaluation scenarios in Sec. \ref{sec:Scenarios}. 

\section{Evaluation}\label{sec:evaluation}
As we implemented all four controllers from Sec.~\ref{sec:control} in Python, we also evaluate them directly in a Python environment, based on the Python toolbox do-mpc \cite{dompc.2017}. In do-mpc, we define the model from Sec.~\ref{sec:model} and use it for the simulation of all four control strategies. In the following, we explain which model parameters, \ac{OTS} scenarios, data, and \ac{KPIs} we use for evaluation.

\subsection{Model Parameters}\label{sec:model_parameters}
The model parameters consist of parameters for the thermal building model and the heat pump. The thermal building model describes the thermal characteristics between heat flows and temperatures in each room. The heat pump model is used to calculate the required electricity to provide heat flows, where the heat pumps' coefficient of performance depends on the outside temperature $T_\text{a}$.

\subsubsection*{Multi-zone thermal building model parameters}
\begin{figure*}[htb]
\centering
    \def\svgwidth{0.65\linewidth}  
    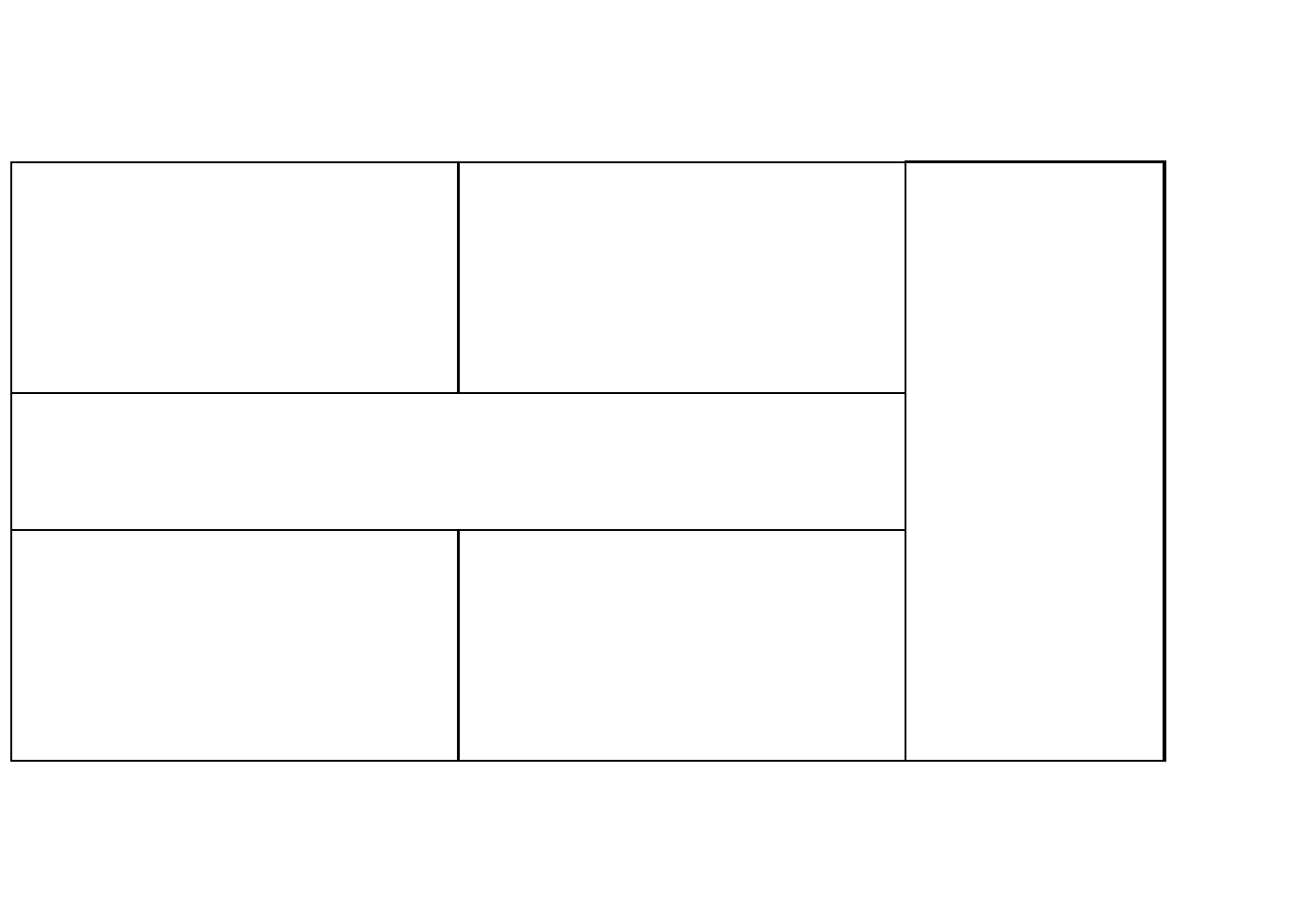
    \caption{Thermal building model for each room  $j$ ($j = 1 \dots 5$)}
    \label{fig:floorplan}
\end{figure*}

First, we define parameter sets by obtaining model parameters from the literature \cite{Madsen.1995} and scaling them from single-zone building level to multi-zone room level (see Fig.~\ref{fig:floorplan}) \cite{Frahm.2022.4}.
We create two different parameter sets, for high and low thermal capacity, to represent different building types. The model parameters used for this evaluation are summarized in Tab.~\ref{tab:model_parameters}.
\begin{table}[htb]
    \centering
    \caption{Model parameters for high and low thermal capacitance}
    \begin{tabularx}{1\linewidth}{lXXXXX}
        \toprule
        par. & $C_{\mathrm{i}}$ & $C_{\mathrm{m}}$ & $R_{\mathrm{a}}$ & $R_{\mathrm{i}}$ & $g_{\mathrm{s}}$ \\
        unit &  \si{\joule \per \kelvin} &  \si{\joule \per \kelvin} & \si{\kelvin \per \watt} & \si{\kelvin \per \watt} & - \\
        \midrule
        high & 3407040  & 11482560  & 0.07345  & 0.001197  &  1.138\\
        low & 1703520  & 5741280  & 0.07345  & 0.001197  &  1.138\\
        \bottomrule 
    \end{tabularx}
    \label{tab:model_parameters}
\end{table}

The parameters for capacity, resistance, and solar heat gain can be defined individually for each room $j$ ($j=1 \dots n$), as illustrated in Fig.~\ref{fig:floorplan} (with $n=5$). In reality, each room has different parameters \footnote{For more information about how to identify individual parameters in multi-zone models, we refer to our previous work \cite{Frahm.2022.3, Frahm.2022.4}.}. For the room-individual evaluation, we multiple the standard parameters from Tab.~\ref{tab:model_parameters} with different random-seed factors for each room. The random seed is once defined and then remains constant over all simulations. The random variable may vary in each room $j$ ($j=1 \dots 5$) by up to $\pm \SI{5}{\percent}$.

\subsubsection*{Heat pump parameters}

We use the model \textit{AERO SLM 3-11 HGL} from the Austrian heat pump manufacturer \textit{iDM Energiesysteme GmbH} \cite{iDMEnergiesystemeGmbH.2020}.
From the manufacturer's technical fact sheet, we obtain the coefficient of performance $\varepsilon_\text{h}$ and the maximum electrical power $P_\text{max}$, as summarized in Tab.~\ref{tab:heatpump_parameters}.
\begin{table}[htb]
    \centering
    \caption{Heat pump parameters for different outside air temperatures $T_\text{a}$}
    \begin{tabularx}{1\linewidth}{llXXXXXXX}
        \toprule
        $T_\text{a}$ in \si{\celsius} & -10 & -7  & 2  & 7 & 10 & 12 & 15 & 20 \\
        \midrule
        $\varepsilon_\text{h}$ &  1.98 & 2.20  & 2.71  & 3.10 & 3.34 & 3.55 & 3.89 & 4.26  \\
        $P_\text{max}$ in \si{ \kilo \watt} & 4.20  & 4.39  & 4.83 & 4.62  & 4.40 & 4.41 & 4.00 & 3.32  \\
        \bottomrule 
    \end{tabularx}
    \label{tab:heatpump_parameters}
\end{table}

Both parameters, $\varepsilon_\text{h}$ and $P_\text{max}$, depend on the outside air temperature $T_\text{a}$ and the heating supply temperature. In this evaluation, we selected a constant heating supply temperature of \SI{55}{\celsius} and a variable outside air temperature, based on the measurements described in Sec.~\ref{sec:data}. Between the discrete values in Tab.~\ref{tab:heatpump_parameters}, we used linear interpolation. 

\subsection{Scenarios for \acf{OTS}}\label{sec:Scenarios}
For the evaluation of the control algorithms, we consider two scenarios, namely \textit{(a) base scenario} and \textit{(b) multi-zone adaptive scenario}. The scenarios differ by their variability of temperature ranges $[T_{\mathrm{lb}_j}(t), T_{\mathrm{ub}_j}(t)]$. The base scenario applies the same ranges for all rooms, while the second allows individual occupancy profiles. 
The temperature ranges are presented in Tab.~\ref{tab:base} and \ref{tab:multi} and applied for each day. Although we assume an office scenario, we do not differentiate between the different days, e.g. between weekdays and weekends, to simplify the evaluation. 

In general, both scenarios apply the \ac{OTS} levels from Tab.~\ref{tab:comfort} in Sec.~\ref{sec:OTS}. Based on the different \ac{OTS} levels, we derive three different control modes, inspired by \cite{Peng.2018}:
\begin{itemize}
    \item \textbf{comfort mode}: \ac{OTS} level I,
    \item \textbf{eco mode}: \ac{OTS} level III,
    \item \textbf{standby mode}: \ac{OTS} level off.
\end{itemize}

While the comfort mode aims for the highest level of \ac{OTS}, the standby mode consumes the fewest energy. The eco mode, however, schedules the reference temperature by approximately $\pm \SI{2}{\kelvin}$ difference compared to the comfort mode (+\SI{1.8}{\kelvin}/-\SI{1.9}{\kelvin} for cooling/heating, see Tab.~\ref{tab:comfort}). 
This eco mode saves energy compared to the comfort mode and also enables fast re-heating/re-cooling compared to the standby mode  \cite{Peng.2018}. In comparison to the comfort mode, the eco mode can save energy in rooms that are less frequently used than office rooms, e.g. bathrooms or kitchens. 

\subsubsection*{(a) Base scenario}
We use two different control modes in the base scenario (see Tab.~\ref{tab:base}): the comfort and standby mode. We apply the comfort mode during working hours from 8{\footnotesize AM} to 5{\footnotesize PM} and the standby mode else. All rooms $j$ ($j = 1 \dots 5$) apply the same modes during the entire evaluation in the base scenario. As a result, the temperature ranges $[T_{\mathrm{lb}_j}(t), T_{\mathrm{ub}_j}(t)]$ in Tab.~\ref{tab:base} are all equal over the different rooms. 
\begin{table}[htb]
    \centering
    \caption{(a) Base scenario \ac{OTS} levels for different periods and rooms}
    \begin{tabularx}{1\linewidth}{lXXXXX}
    \toprule
        Period / room & 1 & 2 & 3 & 4 & 5  \\
        \midrule
        8{\footnotesize AM} to 5{\footnotesize PM} & I & I & I & I & I  \\
        else & off & off & off & off & off  \\
    \bottomrule
    \end{tabularx}
    \label{tab:base}
\end{table}

In contrast, the second scenario uses different modes in different rooms, depending on the use case of each room. 

\subsubsection*{b) Multi-zone adaptive scenario}
As shown in Tab.~\ref{tab:multi}, the temperature ranges $[T_{\mathrm{lb}_j}(t), T_{\mathrm{ub}_j}(t)]$ in all rooms $j$ ($j = 1 \dots 5$) can be different. In this scenario, we also use the eco mode (\ac{OTS} level III) in addition to the comfort (level I) and standby mode (level off).
\begin{table}[htb]
    \centering
    \caption{(b) Multi-zone adaptive scenario \ac{OTS} levels for different periods and rooms}
    \begin{tabularx}{1\linewidth}{lXXXXX}
    \toprule
        Period / room & 1 & 2 & 3 & 4 & 5 \\
        \midrule
        8{\footnotesize AM} to 12{\footnotesize AM}  & I & I & III & III & III  \\
        12{\footnotesize AM} to 1{\footnotesize PM}  & III & III & III & I & III  \\
        1{\footnotesize PM} to 5{\footnotesize PM}  & I & III & III & III & III  \\
        else  & off & off & off & off & off  \\
    \bottomrule
    \end{tabularx}
    \label{tab:multi}
\end{table}

In this multi-zone adaptive scenario from Tab.~\ref{tab:multi}, we let the control operate with a high focus on \ac{OTS} in occupied rooms and energy saving in unoccupied. 
Therefore, we use the comfort mode in the offices (rooms 1 and 2) during working hours and the kitchen (room 4) during lunch breaks from 12{\footnotesize AM} to 1{\footnotesize PM}.  In this scenario, the first office (room 1) is used over the entire working day, except lunch break, and the second office (room 2) only from 8{\footnotesize AM} to 12{\footnotesize AM} (part-time job). The bathroom (room 5) and storage (room 3) should be operated in eco mode during working hours (8{\footnotesize AM} to 5{\footnotesize PM}). 

\subsection{Data}\label{sec:data}

We evaluate the control strategies by using weather data during the winter of 2022/2023, obtained from a weather station in the \textit{KIT EnergyLab} (Karlsruhe, Germany) \cite{Hagenmeyer2016}. For the evaluation, we use the weather measurements of the solar radiation $\dot{q}_\text{s}$ and the ambient temperature $T_\text{a}$) over a period of nine weeks (11/28/2022 -- 02/06/2023). We use the nine weeks of data in time steps of $ \Delta t_k = \SI{15}{\minute}$. 

For the variable electricity tariff $p_\text{buy}$, we use the data of the day-ahead market in Germany from the \textit{EPEX Spot Strombörse}, provided by the aWATTar-API \cite{awattar}. The price is different for every hour of the day and we assume that these prices are directly forwarded to the customers.

\subsection{Metrics}\label{sec:metrics}
We use two \ac{KPIs} to evaluate (i) how accurately a controller meets the desire \ac{OTS} and (ii) how much energy the control strategy therefore consumes. 
Mathematically we define the \ac{KPIs} as the weekly costs $c_\text{m,week}$ in Eq.~\eqref{eq:costs} and mean weekly discomfort $d_\text{m, week}$ in Eq.~\eqref{eq:KPI_discomfort},
\begin{align}
    c_\text{m,week} &=  \sum_{k=1}^{M} \left( p_\text{buy}(t_k) \int_{t_k} P_{\text{el}}(t_k) \, dt_k \right) \label{eq:costs} \\ 
    d_\text{m, week} &= \frac{1}{M} \left(  \sum_{k=1}^{M} \sum_{j=1}^{n} d_{\text{c}_j}(t_k) \right).
    \label{eq:KPI_discomfort}
\end{align}
The \ac{KPIs} consider energy costs and \ac{OTS} during each time-step $k$ for all time steps $M=672$ of each week. The energy costs $c_\text{m,week}$ depend on a dynamic energy tariff $p_\text{buy}(t_k)$ and the consumed electric power $P_{\text{el}}(t_k)$. The discomfort $d_\text{m, week}$ evaluates the discomfort $d_{\text{c}_j}(t_k)$ of the actual room temperature from the allowed \ac{OTS} range. This permitted temperature range is time-variant, depending on room-individual usage/attendance profiles, as introduced in the scenarios in Sec.~\ref{sec:Scenarios}.

Both \ac{KPIs} are competing, which means when one is improved, the other is usually deteriorating. It is the objective to minimize both \ac{KPIs} simultaneously, to have low costs and low discomfort.

\section{Results}\label{sec:results}
In this section, we present the results of the control strategies from Sec.~\ref{sec:control} in the evaluation environment from Sec.~\ref{sec:evaluation}. We compare the different control algorithms in Sec.~\ref{sec:comparison_results}, demonstrate the feasibility of the control strategies in Sec.~\ref{sec:feasibility_results}, discuss the results in Sec.~\ref{sec:discussion}, and show limitations in Sec.~\ref{sec:limitations}.

The results cover the four different control algorithms, \ac{MPC} (ideal and error-free), two variants of \ac{PSC}, and a hysteresis-based two-point controller. The second variant of \ac{PSC} is a simplified version that uses only the price factor (PC, no storage factor) as it is done in \cite{Dengiz.2019}.
The overall results for different parameters and both scenarios over the entire evaluation period of nine weeks can be obtained from Fig.~\ref{fig:results} and \ref{fig:results2}. Fig.~\ref{fig:test} and \ref{fig:test2} illustrate the dynamic response of the thermal building model to the four applied control strategies, exemplarily for the multi-zone adaptive scenario during one week.

\subsection{Comparison of Control Performance}\label{sec:comparison_results}

\begin{figure*}[htb]
     \centering
     \begin{subfigure}[h]{0.49\textwidth}
         \centering
        \includegraphics[width=\textwidth]{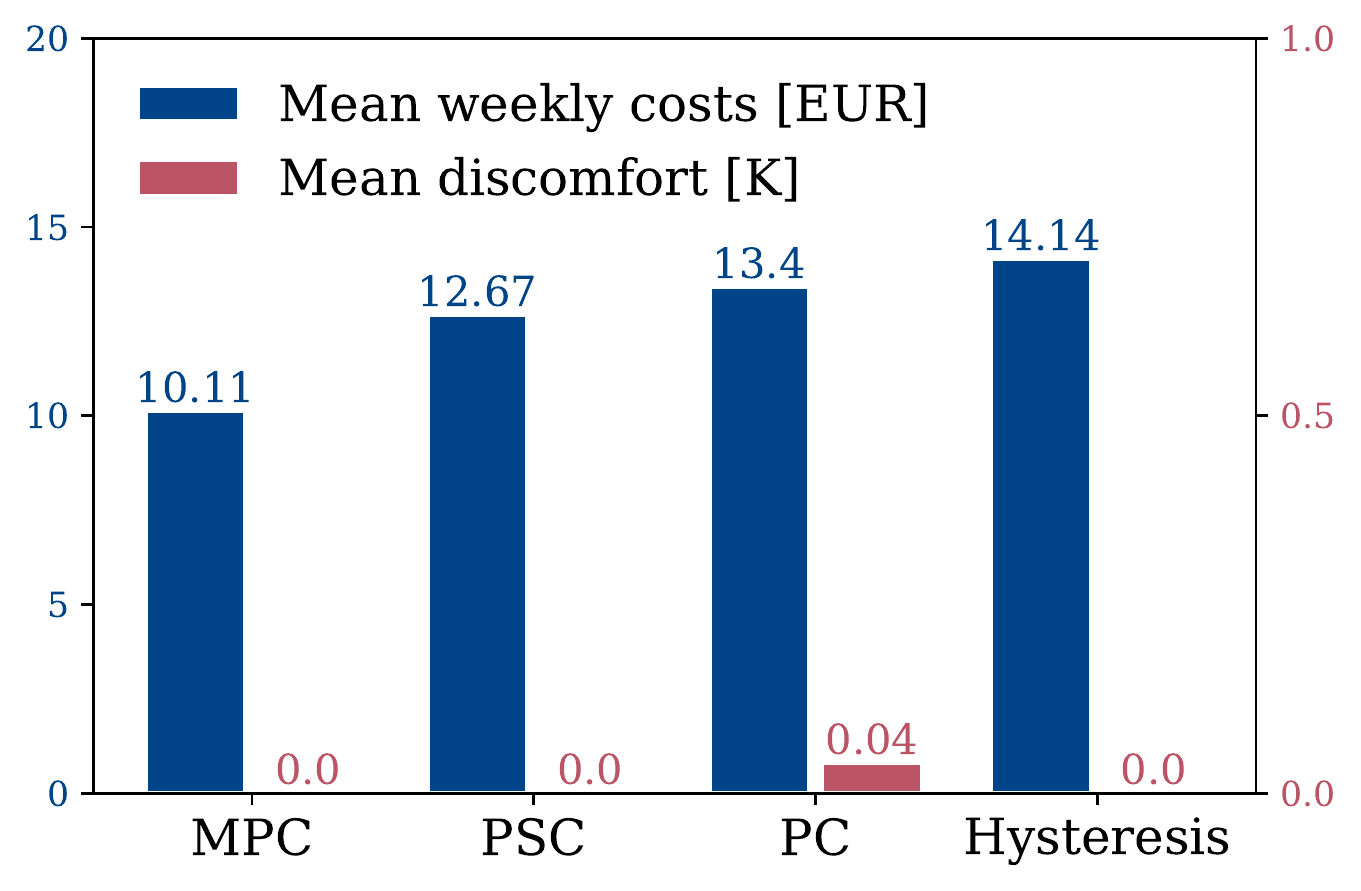}
         \caption{Base scenario, low capacitance}
         \label{fig:base_1}
     \end{subfigure}
     \hfill
     \begin{subfigure}[h]{0.49\textwidth}
         \centering
         \includegraphics[width=\textwidth]{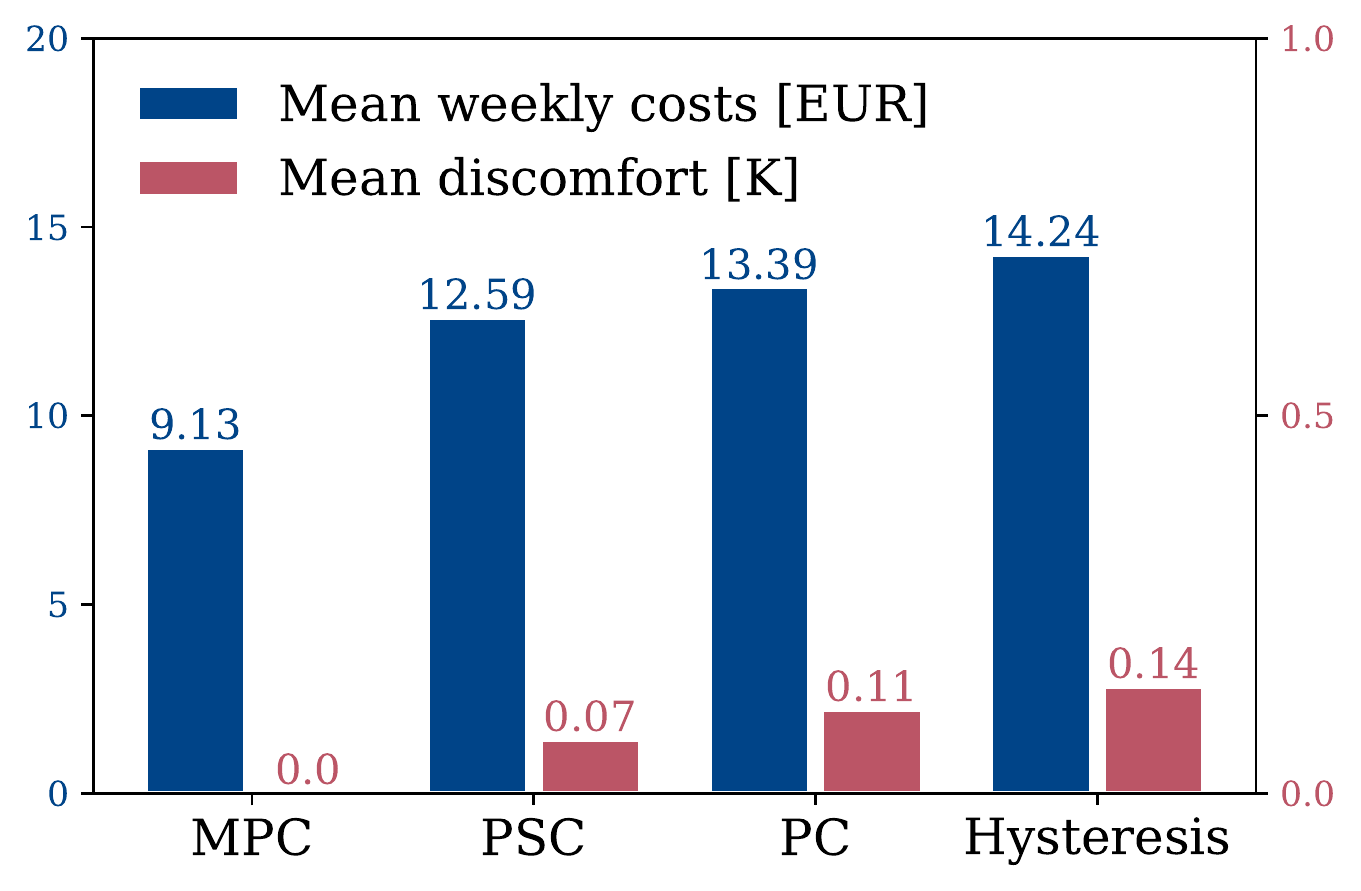}
         \caption{Adaptive scenario, low capacitance}
         \label{fig:multizone_1}
     \end{subfigure}
    \caption{Control results of four controllers, evaluated in two different scenarios with low capacitance}
    \label{fig:results}
\end{figure*}

\begin{figure*}[htb]
     \centering
     \begin{subfigure}[h]{0.49\textwidth}
         \centering
         \includegraphics[width=\textwidth]{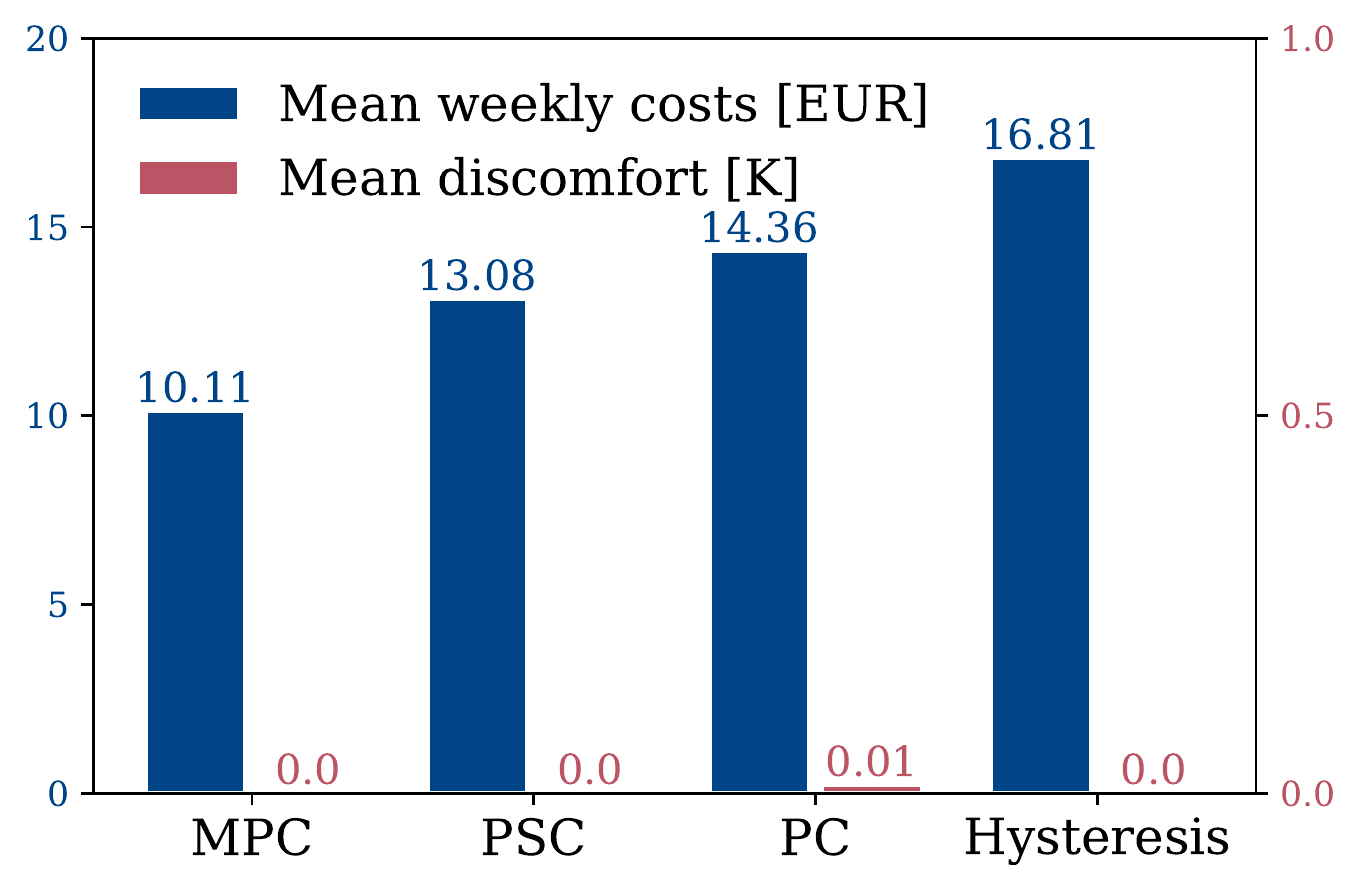}
         \caption{Base scenario, high capacitance}
         \label{fig:base_2}
     \end{subfigure}
     \hfill
     \begin{subfigure}[h]{0.49\textwidth}
         \centering
         \includegraphics[width=\textwidth]{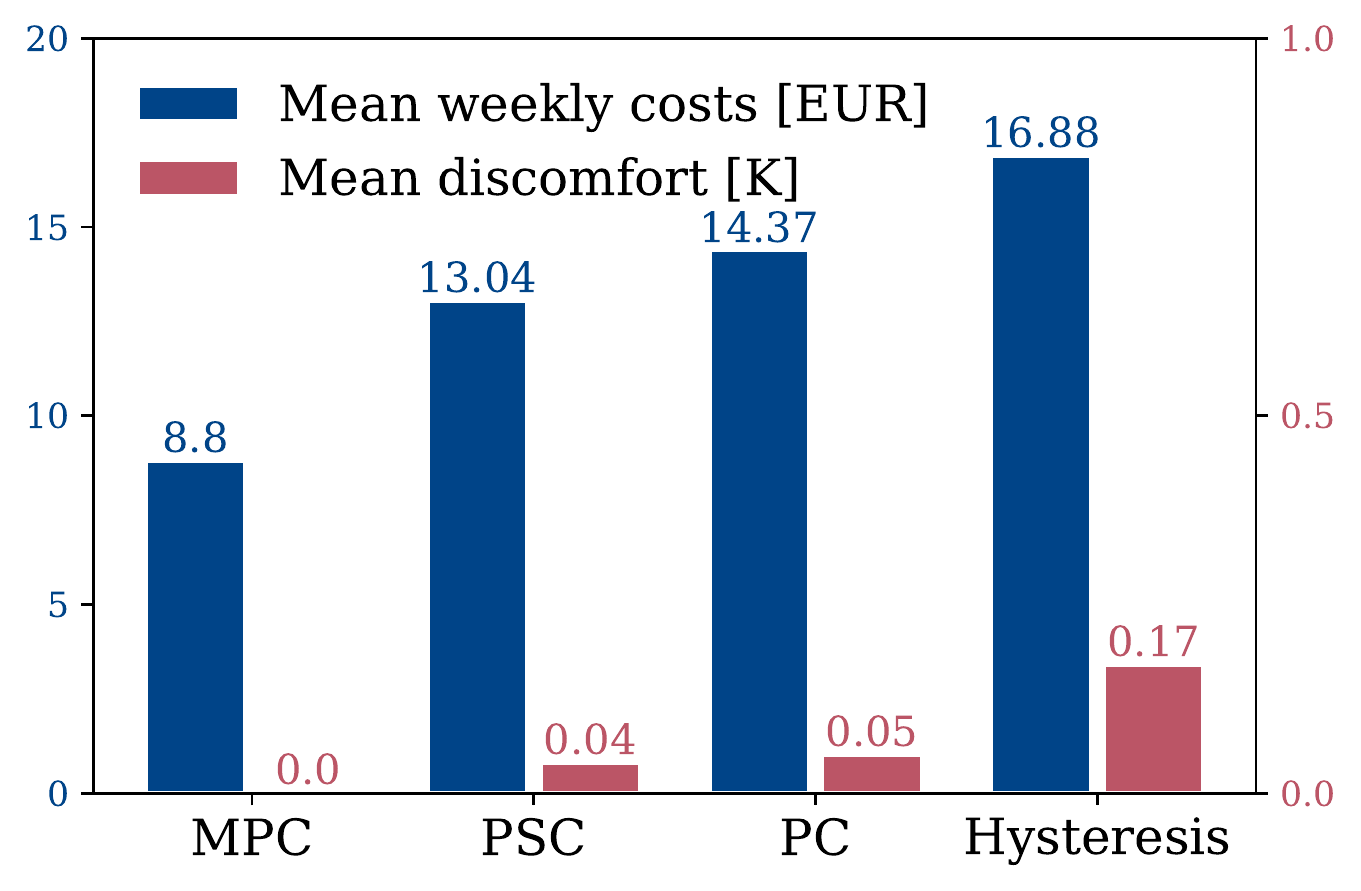}
         \caption{Adaptive scenario, high capacitance}
         \label{fig:multizone_2}
     \end{subfigure}
    \caption{Control results of four controllers, evaluated in two different scenarios with high capacitance}
    \label{fig:results2}
\end{figure*}

We perform evaluations for the four controllers in two scenarios and on two different parameter types over nine different weeks and summarize the results in Fig.~\ref{fig:results} and \ref{fig:results2}.
Fig.~\ref{fig:results} shows the results for low capacitance and Fig.~\ref{fig:results2} for high capacitance. We use Subfig.~(a) for the base case and Subfig.~(b) for the adaptive. 
On the y-axes in Fig.~\ref{fig:results} and \ref{fig:results2}, we visualize the two \ac{KPIs}, the \textit{mean weekly costs} ("costs") and the \textit{mean discomfort} ("discomfort") from Eq.~\eqref{eq:costs} and \eqref{eq:KPI_discomfort}. 

%The results are shown for the two scenarios, the \textit{base scenario}  and the \textit{multi-zone adaptive scenario}, which are based on the temperature ranges in Tab.~\ref{tab:base} and \ref{tab:multi} in Sec.~\ref{sec:Scenarios}, respectively. The parameters are described in Sec.~\ref{sec:model_parameters}.

%%% parameters
\subsubsection*{Parameters}
Comparing Fig.~\ref{fig:results} and \ref{fig:results2} shows the difference between parameters with low and high capacitance. The largest difference can be obtained for the Hysteresis with approximately \SI{15}{\percent} (e.g. in (a) from 16.81 to 14.14). In the following, we will focus on the higher capacitance because it represents larger differences between the different control strategies (see Fig.~\ref{fig:results2}). 

%%% controllers
\subsubsection*{Controllers}
When evaluating the four control strategies in Fig.~\ref{fig:results2}, the \ac{MPC} and \ac{PSC} show superior results in terms of costs and discomfort, compared to the PC and the hysteresis-based two-point controller. In both scenarios, the \ac{MPC} and \ac{PSC} have lower costs than the hysteresis-based two-point controller with more than \SI{39}{\percent} and \SI{22}{\percent} better performance (e.g. in (a) from 16.81 to 10.11 and 13.08). The costs of the simplified PC are between \ac{PSC} and Hysteresis  (\SI{15}{\percent} superior to Hysteresis from 16.81 to 14.36, \SI{9}{\percent} inferior to \ac{PSC} from 14.36 to 13.08).

%%% scenarios
\subsubsection*{Scenarios}
Comparing the different scenarios, the costs of the \ac{MPC} depend more severely on the evaluated scenario, (a) vs. (b), than with the other controllers. 
In the base scenario, the costs of the \ac{MPC} and \ac{PSC} are closer than in the adaptive scenario with \SI{23}{\percent} vs. \SI{33}{\percent} ((a): from 13.08 to 10.11 in vs. (b): 13.04 to 8.8). 
% discomfort
Comparing the adaptive scenario (b) to the base scenario (a), \ac{PSC}, PC, and Hysteresis show low differences in costs and a slightly increased discomfort. This discomfort is more severe with the Hysteresis than with \ac{PSC} or PC (0.17 vs. 0.04 and 0.05). In contrast, the \ac{MPC} has significantly different behavior. On the one hand, the \ac{MPC} violates no comfort (0.0) in both scenarios. On the other hand, the \ac{MPC} reduces costs in the adaptive scenario (b) by \SI{13}{\percent} compared to (a) (from 10.11 to 8.8).

%%% weeks
\subsubsection*{Weeks}
In addition to comparing the four different control strategies, parameters, and scenarios, we present the impact of the different weeks. 
As the \ac{MPC} yields no comfort violations, we compare the costs over all scenarios and weeks in Tab.~\ref{tab:weeks}. As presented in Tab.~\ref{tab:weeks}, the costs of the \ac{MPC} vary from 0 to almost 30 (compare weeks 5 and 3). On average, the \ac{MPC} results in costs of 10 EUR/week.

\begin{table}[htb]
    \centering
    \caption{Influence of the weeks on the optimal solution (\ac{MPC}) with costs in EUR/week}
    \begin{tabularx}{1\linewidth}{Xrrrr}
    \toprule
        week & base/low & base/high & adap./low & adap./high  \\
        \midrule
        1 & 15.15 & 15.15 & 13.89 & 13.33  \\
        2 & 20.23 & 20.19 & 18.46 & 17.67  \\
        3 & 28.82 & 28.89 & 26.58 & 25.31 \\
        4 & 4.35 & 4.23 & 3.72 & 3.47  \\
        5 & 0.08 & 0.01 & 0.00 & 0.00  \\
        6 & 1.54 & 1.51 & 1.12 & 1.00  \\
        7 & 1.08 & 1.00 & 0.80 & 0.66  \\
        8 & 8.8 & 9.24 & 7.88 & 8.56  \\
        9 & 10.93 & 10.77 & 9.75 & 9.17  \\
        average & 10.11 & 10.11 & 9.13 &  8.80 \\
    \bottomrule
    \end{tabularx}
    \label{tab:weeks}
\end{table}

\subsubsection*{Summary}
In summary, we obtain the highest overall performance regarding costs and discomfort with the \ac{MPC} and \ac{PSC}, while the hysteresis-based two-point controller shows the lowest performance. The performance of the simplified PC is between \ac{PSC} and hysteresis. 
The model parameters have the most significant influence on the Hysteresis, while \ac{MPC}, \ac{PSC}, and PC show only a minor influence.
The performance difference between \ac{MPC} and \ac{PSC} depends mostly on the evaluation scenario, where the adaptive scenario is most beneficial for the \ac{MPC}.

In the following, we present insights into the results of the four different control strategies by describing the dynamic behavior from Fig.~\ref{fig:test} and \ref{fig:test2} (see Appendix).

\subsection{Demonstration of Controller Behavior}\label{sec:feasibility_results}
%Control Results for One Week

Fig.~\ref{fig:test} and \ref{fig:test2} illustrate the dynamic behavior of the four controllers, \ac{MPC}, \ac{PSC}, PC, and hysteresis-based two-point controller, on the multi-zone thermal building model with low and high capacitance, respectively. 
First, we generally describe the plots and then show differences between different parameters (low vs. high capacitance) and between the different control strategies.

%% describe plots generally
First, we describe the plots in Fig.~\ref{fig:test} and \ref{fig:test2}. %and the visualized values $T_{\mathrm{i}_j}$, $P_\text{el}$, $T_\text{a}$, and $p$.
The x-axis uses the time in days for one week in December 2022. We visualize week 2 as it represents a relatively high heating demand (see Tab.~\ref{tab:weeks}). 
On the y-axis, we use the air temperatures $T_{\mathrm{i}_j}$ in \si{\celsius} for five rooms ($j=1 \dots 5$) and four control strategies. The blue area shows the permitted temperature ranges for the air temperatures $[T_{\mathrm{lb}_j}(t), T_{\mathrm{ub}_j}(t)]$. The bottom y-axes present the controlled variable $P_\text{el}$, and the time-variable variables $T_\text{a}$, $\dot{q}_\text{s}$, and $p_\text{buy}$.

%% differences from capacitance
Next, we compare the evaluation with low vs. high capacitance (Fig.~\ref{fig:test} vs. \ref{fig:test2}). A lower capacitance shows higher temperature changes from equal heat flows (e.g. see height of temperature peak with \ac{MPC}).
Except for this more sensitive temperature behavior with lower capacitance, the results are relatively similar for both parameters.

%% differences in controllers
Finally, we compare the four different control strategies \ac{MPC}, \ac{PSC}, PC, and Hysteresis. With \ac{MPC}, the temperature is as low as possible while meeting the temperature constraints during all periods. With \ac{MPC}, heating is only applied when the electricity price is at the lowest value, resulting in a cost-optimal \enquote{pre-heating} behavior. Also \ac{PSC} and PC show heating during low electricity prices. However, with \ac{PSC} and PC, the temperature levels are generally higher than with \ac{MPC}. Also, \ac{PSC} reaches lower temperatures than PC, which reduces the amount of used electricity. In contrast to \ac{MPC}, \ac{PSC}, and PC, the Hysteresis shows no explicit reaction to the electricity price. Instead, the Hysteresis has an on/off heating behavior. 

Overall, the results from Fig.~\ref{fig:test} vs. \ref{fig:test2} illustrated the feasibility to control a multi-zone thermal building environment with differently complex control strategies. The different controllers showed different heating behaviors and resulting temperatures. 

\subsection{Discussion}\label{sec:discussion}
In this section, we discuss the results from Sec.~\ref{sec:comparison_results} and \ref{sec:feasibility_results} and the differences between parameters, scenarios, controllers, and weeks. 
Depending on the controller's complexity, the decision-making process of the controllers can include variables such as room-individual temperature limits, the electricity price, weather conditions, or room-individual thermal building dynamics (from a building model). In the following, we further discuss the results.

\subsubsection*{Parameters}
The higher thermal sensitivity towards heat flows $\dot{Q}$ of the evaluation environment with lower thermal capacitance $C$ results from the inverse proportionality between temperature changes $\Delta T$ and heat capacitance:  $\Delta T = \frac{1}{C} \cdot \dot{Q}$ (see Eq~\eqref{eq:capacitor}). As a result, with more heat capacitance more internal energy (heat) can be stored. 

Although all control strategies generally exploit the thermal capacitance, only \ac{MPC} explicitly uses a model for it in its decision-making process. The impact of the different control strategies is discussed in the following.

\subsubsection*{Controllers}
The four controllers differ in their complexity and how much knowledge about future system behavior they require. Depending on the controller's complexity, the performance varies in the present study. 
The control performance of the \ac{PSC} is superior to the hysteresis-based two-point controller and inferior to the ideal \ac{MPC}. The simplified PC performs between \ac{PSC} and Hysteresis. 

%%% explanation of results
% hysteresis
The hysteresis-based two-point controller uses only minimal and maximal temperatures $[T_{\mathrm{lb}_j}(t), T_{\mathrm{ub}_j}(t)]$ without any forecasts or models. 
When a maximal temperature is reached, it heats over a defined period. As no explicit knowledge about the energy price can be considered, this control strategy is relatively cost-ineffective.

%PSC
The \ac{PSC} requires knowledge about the temperature ranges, but also about the energy tariff and the heat pump modulation. 
It tries to meet a reference temperature that is in the middle of the minimal and maximal ranges, while also considering periods of low energy prices.
Exploiting this knowledge reduces the energy costs of the \ac{PSC} compared to the hysteresis-based two-point controller because the \ac{PSC} can apply heating during periods of low energy prices. 
Compared to \ac{PSC}, the simplified PC heats more conservatively as no storage factor is applied. 

% MPC
The \ac{MPC} uses the largest amount of available information, which increases its performance accordingly.
It does not only use temperature ranges, energy tariffs, and heat pump modulation. In addition, the \ac{MPC} needs a thermal building model and weather forecasts. With that internal control model and the forecasts, the \ac{MPC} can predict future system behavior in advance and schedule the heating load optimally. 
As a result, the \ac{MPC} applies the lowest possible temperatures to save energy and pre-heats in advance during the lowest energy prices to additionally save costs.

\subsubsection*{Scenarios}
The most striking result in the evaluation of the scenarios, (a) base vs. (b) multi-zone adaptive scenario, was obtained with the \ac{MPC}. While \ac{PSC}, PC, and Hysteresis yielded only insignificant cost differences between both scenarios, the \ac{MPC} reduced costs by up to \SI{13}{\percent}.  
%As a results, the \ac{MPC} outperformed the second best controller, \ac{PSC} by \SI{33}{\percent} in scenario (b) and \SI{23}{\percent} in (a).
The results indicate the potential of room-individual building control with relaxed constraints based on individual occupancy presence. 
During unoccupied periods, the \ac{MPC} could save energy by reducing the temperature to its full potential.

\subsubsection*{Weeks}
The costs for heating vary significantly based on the evaluated week, from 0 to almost 30. The weeks differ not only by the weather conditions and the resulting heating demand but also by the electricity price. As single weeks cannot yield sufficiently representative results, averaging over multiple weeks is required for meaningful results.

\subsubsection*{Summary}
Overall, the results demonstrated the importance and feasibility of multi-zone building control. We presented differently complex control strategies that can be applied depending on the availability of temperature ranges, forecasts, and models. With more information available, the control performance can be increased. Even when no models are available, adequate performance can be obtained with our proposed \ac{PSC} control strategy. The four developed control strategies are applicable to a variety of buildings as we presented different evaluation parameters with high and low capacitance and for two different scenarios with different temperature constraints. 

The controllers perform as expected where a higher complexity and use of more information improve the control quality. 
While the \ac{PSC} outperforms the hysteresis-based two-point controller and simplified PC, \ac{PSC} is inferior to an ideal \ac{MPC}. 
On the one hand, the \ac{MPC} showed potential for cost optimization in thermal building control, especially in the multi-zone adaptive scenario. 
On the other hand, the \ac{MPC} is significantly more complex to design than other strategies, requiring a thermal model for each room and a forecast, which we both assumed to be error-free for our case study.

\subsection{Limitations}\label{sec:limitations}
The evaluation of control strategies in this work is based on simulation results, which can neglect several effects from the real application. The control strategies are performed on a multi-zone thermal building model instead of a real building. The model parameters are based on literature values instead of identification from parameter identification. The model and weather forecasts of the \ac{MPC} are assumed as error-free. 

The evaluation is limited to a heating scenario, where weather data is used for nine weeks during winter in Karlsruhe, Germany. A cooling scenario is not investigated because the cooling demand in Germany is lower than in other regions of the world. 
The evaluated scenarios consider no self-produced \ac{PV}, battery, appliances, or thermal water storage in the optimization.
\section{Conclusion}
\label{sec:conclusion}
\acresetall

In this study, we investigate how a novel multi-zone \ac{PSC} can provide \ac{DR} while considering room-individual \ac{OTS} without using a thermal building model and weather forecasts. Therefore, we develop four different control strategies, a multi-zone evaluation environment with different thermal parameters for high and low thermal capacitance, and two different scenarios to compare the controllers. We compare the \ac{PSC} with an ideal, error-free \ac{MPC}, with a simplified variant without a storage factor (\acs{PC}), and hysteresis-based two-point controller as upper and lower benchmarks, respectively.

The ideal \ac{MPC} and \ac{PSC} achieve higher control performance than the hysteresis-based two-point controller in terms of energy costs and mean discomfort in all scenarios and for all parameters. With high capacitance, the \ac{PSC} leads to a cost reduction of \SI{22}{\percent}, \acs{PC} \SI{15}{\percent}, while the \ac{MPC} achieves improvements of more than \SI{39}{\percent}. Under consideration that the \ac{PSC} requires no models and no forecasts, this control strategy seems especially beneficial for real-world control applications. 
Our developed control approach is easy to implement and can be used for every building without large-scale adjustments. Further, it can include other external signals in its decision-making like the load of the electricity grid or a generation signal of renewable energy sources. Thus, it can contribute to balancing electricity demand and supply and lead to better utilization of renewable energy sources in future energy systems. 

In future work, we want to apply the developed control strategy to a real-world application. For the \ac{MPC} real-world application, we need to perform parameter identification and design a state estimator. 
For a more realistic scenario, we plan to include more relevant components into the optimization, e.g. thermal water storage, \ac{PV} self-production and -consumption, and batteries. 
\section*{Data Availability}

We added the following supplementary materials to an
open-source online repository on \href{https://github.com/Occupant-Oriented-Demand-Response/Conrol-Results}{GitHub}:

\begin{itemize} 
\item results of the four control strategies for all individual weeks in both scenarios and for both parameters,
\item used input data for the electricity price and weather data,
\item commented Python code of the four control strategies.
\end{itemize}

\href{https://github.com/Occupant-Oriented-Demand-Response/Conrol-Results}{https://github.com/Occupant-Oriented-Demand-Response/Conrol-Results}.
\section*{Acknowledgment}

This work was conducted within the project FlexKälte, funded by the German Federal Ministry for Economic Affairs and Climate Action (BMWK), and supported by the Helmholtz Association under the program Energy System Design (ESD).
The authors would like to thank their colleagues from the Energy Lab and the Institute for Automation and Applied Informatics (IAI) for all the fruitful discussions and collaborations.
\appendix \label{sec:appendix}

\section{Control Results for One Week}
\begin{figure*}[htb]
    \centering
    \includegraphics[width=\linewidth]{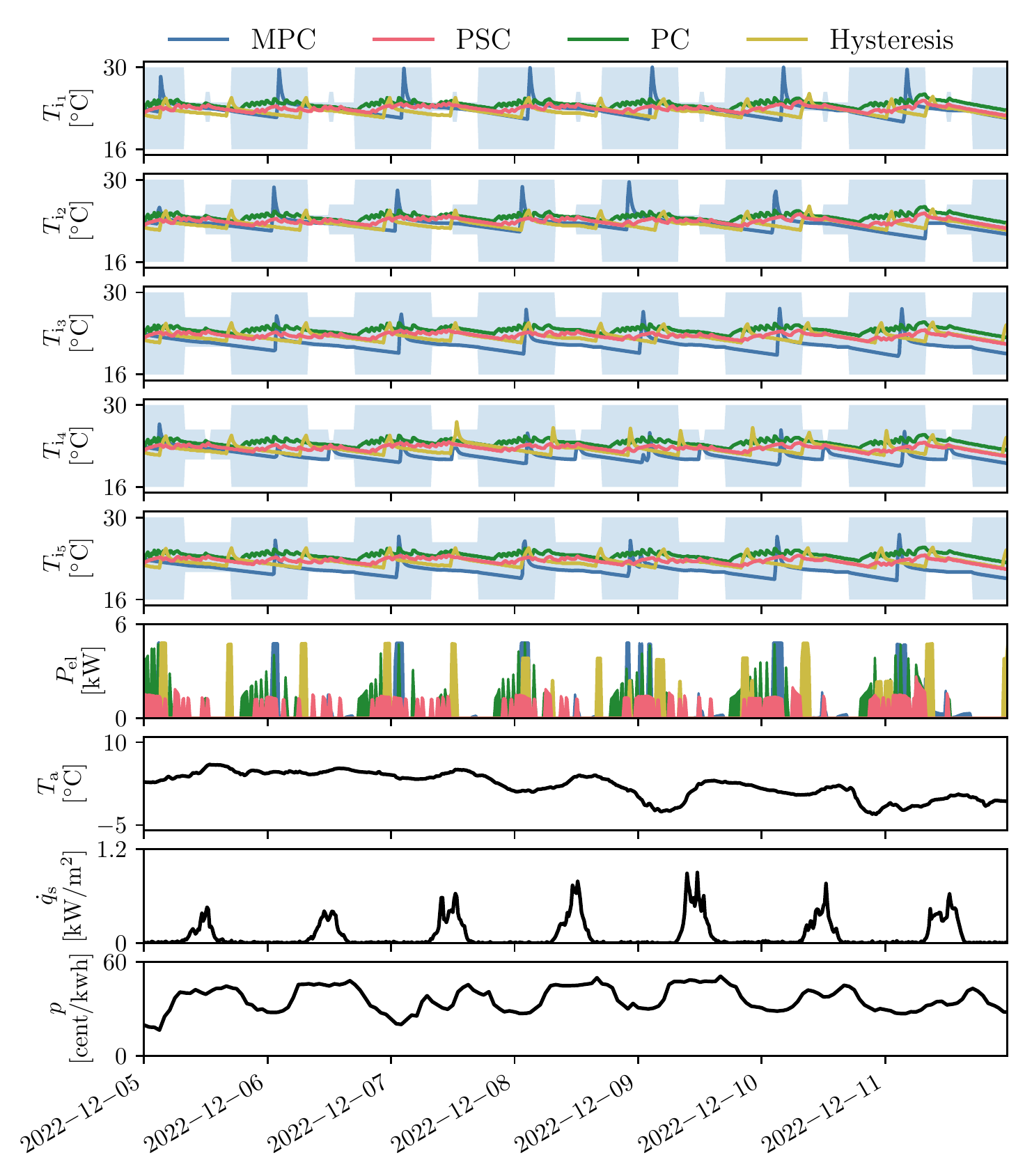}
    \caption{Control results in the adaptive scenario with low capacitance for week 2}
    \label{fig:test}
\end{figure*}

\begin{figure*}[htb]
    \centering
    \includegraphics[width=\linewidth]{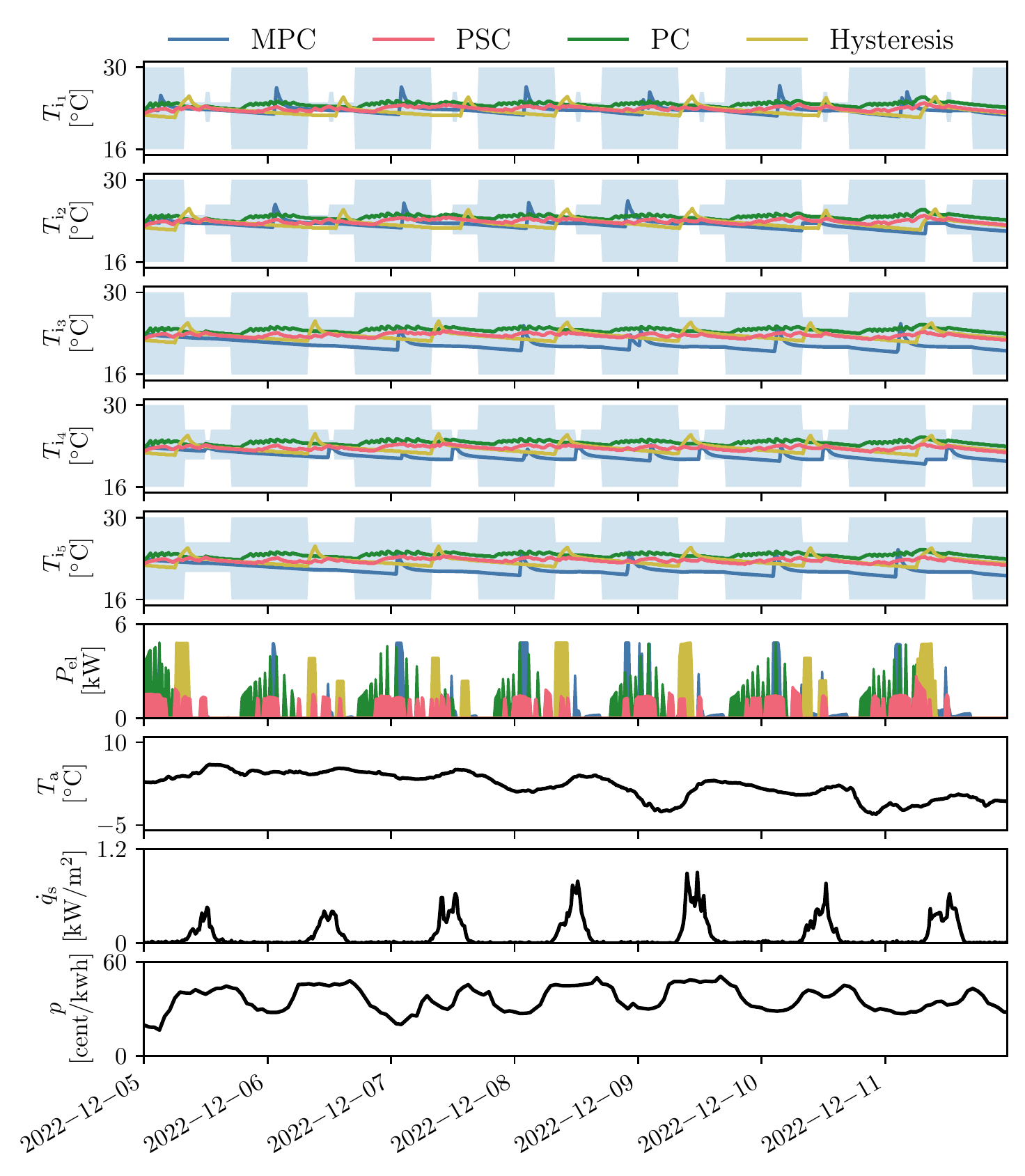}
    \caption{Control results in the adaptive scenario with high capacitance for week 2}
    \label{fig:test2}
\end{figure*}

The control results for one week, as described in Sec.~\ref{sec:feasibility_results}, are presented in Fig.~\ref{fig:test} and \ref{fig:test2}.

 \bibliographystyle{elsarticle-num} 
 \bibliography{Sources}





% \end{thebibliography}
\end{document}